\documentclass[conference]{IEEEtran}

\usepackage{graphicx}
\usepackage[utf8]{inputenc}
\usepackage{amsmath}
\usepackage{array}
\usepackage[hyphens]{url}
\usepackage[table]{xcolor}
\usepackage{xspace}
\usepackage{pbox}
\PassOptionsToPackage{hyphens}{url}\usepackage{url}
\usepackage{letltxmacro}
\usepackage{cite} 
\usepackage{ifthenx}
\usepackage{collcell}
\usepackage{nicefrac}
\usepackage{comment}
\usepackage[all]{nowidow}
\usepackage{filecontents}
\usepackage{listings}
\usepackage{algorithmic}
\usepackage[ruled]{algorithm2e}
\usepackage{lipsum}
\reversemarginpar
\usepackage{todonotes}
\usepackage{marginnote} 

\usepackage{makecell}
\usepackage{paralist}
\usepackage{flushend}
\usepackage{ifthenx}
\usepackage{xspace}
\usepackage{soul} 
\usepackage{csvenhanced}
\usepackage{etex}
\usepackage{bytefield}
\usepackage{amsfonts}
\usepackage{tcolorbox}
\usepackage{arydshln}
\usepackage{siunitx}
\usepackage{booktabs}
\usepackage{color}
\usepackage[caption=false,font=footnotesize]{subfig}
\usepackage{multirow}
\usepackage{hhline}
\usepackage{float}
\usepackage[inline]{enumitem}
\usepackage{rotating}
\usepackage{bytefield}

\usepackage{tikz}
\usepackage{pgfplots}
\usetikzlibrary{pgfplots.groupplots}
\usetikzlibrary{arrows}
\usetikzlibrary{patterns}
\usetikzlibrary{positioning}
\usetikzlibrary{decorations.pathreplacing}
\usetikzlibrary{shapes.arrows}
\usetikzlibrary{pgfplots.groupplots}
\pgfplotsset{compat=newest}

\LetLtxMacro{\oldtodo}{\todo}
\renewcommand{\todo}[2][]{\oldtodo[fancyline,size=\footnotesize,#1]{#2}}
\renewcommand{\todo}[1]{\oldtodo[fancyline,size=\footnotesize]{#1}}

\renewcommand{\subsubsection}[1]{\textbf{#1}.}

\DeclareMathOperator{\KB}{KB}

\newcommand{\xor}{\oplus}

\newcommand{\SIx}[1]{\SI{#1}\relax}

\newcommand{\etal}{et~al.\ } 
\newcommand{\ie}{\textit{i.e.},\ } 
\newcommand{\eg}{e.g.,\ } 

\newcommand{\FlushReload}{\emph{Flush+\allowbreak Reload}\xspace}

\newcommand{\PrimeProbe}{\emph{Prime+\allowbreak Probe}\xspace}
\newcommand{\SGXone}{SGX1\xspace}
\newcommand{\SGXtwo}{SGX2\xspace}

\newcommand{\mbedTLS}{\emph{mbedTLS}\xspace}

\definecolor{corange}{HTML}{F36523}
\definecolor{lorange}{HTML}{FFC20F}
\definecolor{cgreen}{HTML}{8EC63F}
\definecolor{cpurple}{HTML}{942977}
\definecolor{cblue}{HTML}{27999F}
\definecolor{lblue}{HTML}{B2B2FF}
\definecolor{lred}{HTML}{FFB2B2}
\definecolor{llgray}{HTML}{CCCCCC}
\definecolor{lgray}{HTML}{AAAAAA}

\newcommand{\zero}{`0'\xspace}
\newcommand{\zeros}{`0's\xspace}
\newcommand{\one}{`1'\xspace}
\newcommand{\ones}{`1's\xspace}

\usepackage{pifont}

\lstdefinelanguage
   [x64]{Assembler}     
   [x86masm]{Assembler} 
   {morekeywords={CDQE,CQO,CMPSQ,CMPXCHG16B,JRCXZ,LODSQ,MOVSXD, %
                  POPFQ,PUSHFQ,SCASQ,STOSQ,IRETQ,RDTSCP,SWAPGS, %
                  rax,rdx,rcx,rbx,rsi,rdi,rsp,rbp, %
                  r8,r8d,r8w,r8b,r9,r9d,r9w,r9b}} 

\newfloat{listing}{tbhp}{lst}
\floatname{listing}{Listing}

\begin{document}
\title{Malware Guard Extension:\\Using SGX to Conceal Cache Attacks\\(Extended Version)}

\author{\IEEEauthorblockN{Michael Schwarz}
\IEEEauthorblockA{Graz University of Technology\\
Email: michael.schwarz@iaik.tugraz.at}
\and
\IEEEauthorblockN{Samuel Weiser}
\IEEEauthorblockA{Graz University of Technology\\
Email: samuel.weiser@iaik.tugraz.at}
\and
\IEEEauthorblockN{Daniel Gruss}
\IEEEauthorblockA{Graz University of Technology\\
Email: daniel.gruss@iaik.tugraz.at}
\and
\IEEEauthorblockN{Clémentine Maurice}
\IEEEauthorblockA{Graz University of Technology\\
Email: clementine.maurice@iaik.tugraz.at}
\and
\IEEEauthorblockN{Stefan Mangard}
\IEEEauthorblockA{Graz University of Technology\\
Email: stefan.mangard@iaik.tugraz.at}}

%

\maketitle

\begin{abstract}
In modern computer systems, user processes are isolated from each other by the operating system and the hardware.
Additionally, in a cloud scenario it is crucial that the hypervisor isolates tenants from other tenants that are co-located on the same physical machine. 
However, the hypervisor does not protect tenants against the cloud provider and thus the supplied operating system and hardware.
Intel SGX provides a mechanism that addresses this scenario.
It aims at protecting user-level software from attacks from other processes, the operating system, and even physical attackers.

In this paper, we demonstrate fine-grained software-based side-channel attacks from a malicious SGX enclave targeting co-located enclaves. 
Our attack is the first malware running on real SGX hardware, abusing SGX protection features to conceal itself.
Furthermore, we demonstrate our attack both in a native environment and across multiple Docker containers. 
We perform a \PrimeProbe cache side-channel attack on a co-located SGX enclave running an up-to-date RSA implementation that uses a constant-time multiplication primitive.
The attack works although in SGX enclaves there are no timers, no large pages, no physical addresses, and no shared memory.
In a semi-synchronous attack, we extract 96\,\% of an RSA private key from a single trace. 
We extract the full RSA private key in an automated attack from 11 traces within 5 minutes.

\end{abstract}

%
\IEEEpeerreviewmaketitle

%
\section{Introduction} \label{sec:introduction}
%
Modern operating systems isolate user processes from each other to protect secrets in different processes.
Such secrets include passwords stored in password managers or private keys to access company networks.
Leakage of these secrets can compromise both private and corporate systems.
Similar problems arise in the cloud.
Therefore, cloud providers use virtualization as an additional protection using a hypervisor.
The hypervisor isolates different tenants that are co-located on the same physical machine.
However, the hypervisor does not protect tenants against a possibly malicious cloud provider.

Although hypervisors provide functional isolation, side-channel attacks are often not considered.
Consequently, researchers have demonstrated various side-channel attacks, especially those exploiting the cache~\cite{Ge2016}.
Cache side-channel attacks can recover cryptographic secrets, such as AES~\cite{Apecechea2014a,Guelmezoglu2015} and RSA~\cite{Inci2015} keys, across virtual machine boundaries.

Intel introduced a new hardware extension SGX (Software Guard Extensions)~\cite{Intel_SGX} in their CPUs, starting with the Skylake microarchitecture.
SGX is an isolation mechanism, aiming at protecting code and data from modification or disclosure even if all privileged software is malicious~\cite{Costan2016}. 
This protection uses special execution environments, so-called enclaves, which work on memory areas that are isolated from the operating system by the hardware. 
The memory area used by the enclaves is encrypted to protect the application's secrets from hardware attackers.
Typical use cases include password input, password managers, and cryptographic operations.
Intel recommends storing cryptographic keys inside enclaves and claims that side-channel attacks ``are thwarted since the memory is protected by hardware encryption''~\cite{Intel_HardeningSGX}.

Apart from protecting software, the hardware-supported isolation led to fear of super malware inside enclaves.
Rutkowska~\cite{Rutkowska2013} outlined a scenario where a benign-looking enclave fetches encrypted malware from an external server and decrypts and executes it within the enlave. 
In this scenario, it is impossible to debug, reverse engineer, or in any other way analyze the executed malware. 
Aumasson~\etal\cite{Aumasson2016} and Costan~\etal\cite{Costan2016} eliminated this fear by arguing that enclaves always run with user space privileges and can neither issue syscalls nor perform any I/O operations. 
Moreover, SGX is a highly restrictive environment for implementing cache side-channel attacks. 
Both state-of-the-art malware and side-channel attacks rely on several primitives that are not available in SGX enclaves.
Consequently, no enclave malware has been demonstrated on real hardware so far.

In this paper, we show that it is very well possible for enclave malware to attack its hosting system.
We demonstrate a cache attack from within a malicious enclave that is extracting secret keys from co-located enclaves.
Our proof-of-concept malware is able to recover RSA keys by monitoring cache access patterns of an RSA signature process in a semi-synchronous attack. 
The malware code is completely invisible to the operating system and cannot be analyzed due to the isolation provided by SGX. 
In an even stronger attack scenario, we show that an additional isolation using Docker containers does not protect against this kind of attack.

We make the following contributions:

\begin{compactenum}
 \item We demonstrate that, despite the restrictions of SGX, cache attacks can be performed from within an enclave to attack a co-located enclave.
 \item By combining DRAM and cache side channels, we present a novel approach to recover physical address bits even if \SI{2}{\mega B} pages are unavailable. 
 \item We show that it is possible to have highly accurate timings within an enclave without access to the native timestamp counter, which is even more accurate than the native one. 
 \item
We demonstrate a fully automated end-to-end attack on the RSA implementation of the wide-spread \mbedTLS library. 
We extract \SI{96}{\percent} of an RSA private key from a single trace and the full key from 11 traces within 5 minutes.
\end{compactenum}

Section~\ref{sec:background} presents the background required for our work.
Section~\ref{sec:threat} outlines the threat model and our attack scenario.
Section~\ref{sec:pp} describes the measurement methods and the online phase of the malware.
Section~\ref{sec:keyextraction} explains the key recovery techniques used in the offline phase.
Section~\ref{sec:evaluation} evaluates the attack against an up-to-date RSA implementation.
Section~\ref{sec:countermeasures} discusses several countermeasures. 
Section~\ref{sec:conclusion} concludes our work.

%
\section{Background} \label{sec:background}
%

\subsection{Intel SGX in Native and Virtualized Environments}\label{sec:sgx}
Intel Software Guard Extensions~(SGX) are a new set of x86 instructions introduced with the Skylake microarchitecture. SGX allows protecting the execution of user programs in so-called enclaves. Only the enclave can access its own memory region, any other access to it is blocked by the CPU. As SGX enforces this policy in hardware, enclaves do not need to rely on the security of the operating system. In fact, with SGX the operating system is generally not trusted. By doing sensitive computation inside an enclave, one can effectively protect against traditional malware, even if such malware has obtained kernel privileges. Furthermore, it allows running secret code in a cloud environment without trusting the cloud provider's hardware and operating system.

An enclave resides in the virtual memory area of an ordinary application process. 
When creating an enclave, a virtual memory region is reserved for the enclave. 
This virtual memory region can only be backed by physically protected pages from the so-called Enclave Page Cache~(EPC). 
In SGX, the operating system is in charge of mapping EPC pages correctly. 
However, any invalid or malicious page mapping is detected by the CPU to maintain enclave protection. 
The EPC itself is a contiguous physical block of memory in DRAM that is transparently encrypted using a dedicated hardware encryption module. 
This protects enclaves against hardware attacks trying to read or manipulate enclave content in DRAM. 

Creation and loading of enclaves are done by the operating system. To protect the integrity of enclave code, the loading procedure is measured by the CPU. If the resulting measurement does not match the value specified by the enclave developer, the CPU will refuse to run the enclave. During enclave loading, the operating system has full access to the enclave binary. At this point anti-virus scanners can hook in to analyze the enclave binary before it is executed. Enclave malware will attempt to hide from anti-virus scanners by encrypting malicious payload.

Since enclave code is known to the (untrusted) operating system, it cannot carry hard-coded secrets. Any secret information might only be provisioned to the enclave during runtime. Before giving secrets to an enclave, a provisioning party has to ensure that the enclave has not been tampered with. SGX therefore provides remote attestation, which proves correct enclave loading via the aforementioned enclave measurement.

SGX comes in two versions. \SGXone specifies basic enclave operation. Moreover, all enclave memory pages have to be allocated at enclave creation. To account for limited memory resources, enclave pages can be swapped out and in at runtime. \SGXtwo extends SGX with dynamic memory management, allowing to allocate new enclave pages at runtime. However, we do not use \SGXtwo features and thus presume that our attack is applicable to \SGXtwo as well.

At the time of writing, no hypervisor with SGX support was available to us. 
However, Docker~\cite{Docker2016} has support for Intel's SGX.
Docker is an operating-system-level virtualization software that allows applications with all their dependencies to be packed into one container.
It has emerged as a standard runtime for containers on Linux and can be used on multiple cloud providers. 
Unlike virtual machines, Docker containers share the kernel and other resources with the host system, requiring fewer resources than a virtual machine. 
Docker isolates processes from each other but does not give a full isolation guarantee such as virtual machines. 
Arnautov~\etal\cite{Arnautov2016} proposed to combine Docker containers with SGX to create secure containers. 

\subsection{Microarchitectural Attacks}
Microarchitectural attacks exploit hardware properties that allow inferring information on other processes running on the same system.
In particular, cache attacks exploit the timing difference between the CPU cache and the main memory.
They have been the most studied microarchitectural attacks for the past 20 years, and were found to be powerful attacks able to derive cryptographic secrets~\cite{Kocher1996,Page2002,Bernstein2005,Percival2005}.

While early attacks focused on the L1 caches, more modern attacks target the last-level cache, which is shared among all CPU cores. Last-level caches (LLC) are usually built as $n$-way set-associative caches. 
They consist of $S$ cache sets and each cache set consists of $n$ cache ways with a size of \SI{64}{B}. 
The physical address determines to which cache set and byte offset a variable maps. 
The lowest $6$ bits determine the byte offset within a cache way, the following $\log_2 S$ bits starting with bit $6$ determine the cache set. 
Only the cache way is not derived from the physical address but chosen by the CPU using its cache replacement policy. 

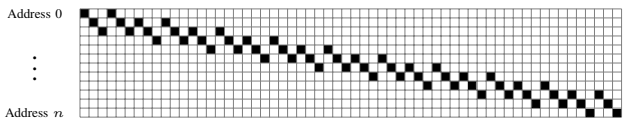
\begin{figure}
\centering
 \begin{tikzpicture}[scale=0.6]
\draw[step=0.2cm,gray,very thin] (0,0.2) grid (12,2.6);
\node at (-1,2.5) {\tiny{Address 0}};
\node at (-1, 1.45) {$\vdots$};
\node at (-1,0.3) {\tiny{Address $n$}};
\filldraw[fill=black, draw=black] (0.02, 2.58) rectangle (0.18, 2.42);
\filldraw[fill=black, draw=black] (0.62, 2.58) rectangle (0.78, 2.42);
\filldraw[fill=black, draw=black] (0.22, 2.38) rectangle (0.38, 2.22);
\filldraw[fill=black, draw=black] (0.82, 2.38) rectangle (0.98, 2.22);
\filldraw[fill=black, draw=black] (1.22, 2.38) rectangle (1.38, 2.22);
\filldraw[fill=black, draw=black] (1.82, 2.38) rectangle (1.98, 2.22);
\filldraw[fill=black, draw=black] (0.42, 2.18) rectangle (0.58, 2.02);
\filldraw[fill=black, draw=black] (1.02, 2.18) rectangle (1.18, 2.02);
\filldraw[fill=black, draw=black] (1.42, 2.18) rectangle (1.58, 2.02);
\filldraw[fill=black, draw=black] (2.02, 2.18) rectangle (2.18, 2.02);
\filldraw[fill=black, draw=black] (2.42, 2.18) rectangle (2.58, 2.02);
\filldraw[fill=black, draw=black] (3.02, 2.18) rectangle (3.18, 2.02);
\filldraw[fill=black, draw=black] (1.62, 1.98) rectangle (1.78, 1.82);
\filldraw[fill=black, draw=black] (2.22, 1.98) rectangle (2.38, 1.82);
\filldraw[fill=black, draw=black] (2.62, 1.98) rectangle (2.78, 1.82);
\filldraw[fill=black, draw=black] (3.22, 1.98) rectangle (3.38, 1.82);
\filldraw[fill=black, draw=black] (3.62, 1.98) rectangle (3.78, 1.82);
\filldraw[fill=black, draw=black] (4.22, 1.98) rectangle (4.38, 1.82);
\filldraw[fill=black, draw=black] (2.82, 1.78) rectangle (2.98, 1.62);
\filldraw[fill=black, draw=black] (3.42, 1.78) rectangle (3.58, 1.62);
\filldraw[fill=black, draw=black] (3.82, 1.78) rectangle (3.98, 1.62);
\filldraw[fill=black, draw=black] (4.42, 1.78) rectangle (4.58, 1.62);
\filldraw[fill=black, draw=black] (4.82, 1.78) rectangle (4.98, 1.62);
\filldraw[fill=black, draw=black] (5.42, 1.78) rectangle (5.58, 1.62);
\filldraw[fill=black, draw=black] (4.01, 1.58) rectangle (4.18, 1.42);
\filldraw[fill=black, draw=black] (4.62, 1.58) rectangle (4.78, 1.42);
\filldraw[fill=black, draw=black] (5.01, 1.58) rectangle (5.18, 1.42);
\filldraw[fill=black, draw=black] (5.62, 1.58) rectangle (5.78, 1.42);
\filldraw[fill=black, draw=black] (6.02, 1.58) rectangle (6.18, 1.42);
\filldraw[fill=black, draw=black] (6.62, 1.58) rectangle (6.78, 1.42);
\filldraw[fill=black, draw=black] (5.22, 1.38) rectangle (5.38, 1.22);
\filldraw[fill=black, draw=black] (5.82, 1.38) rectangle (5.98, 1.22);
\filldraw[fill=black, draw=black] (6.22, 1.38) rectangle (6.38, 1.22);
\filldraw[fill=black, draw=black] (6.82, 1.38) rectangle (6.98, 1.22);
\filldraw[fill=black, draw=black] (7.22, 1.38) rectangle (7.38, 1.22);
\filldraw[fill=black, draw=black] (7.82, 1.38) rectangle (7.98, 1.22);
\filldraw[fill=black, draw=black] (6.42, 1.18) rectangle (6.58, 1.02);
\filldraw[fill=black, draw=black] (7.02, 1.18) rectangle (7.18, 1.02);
\filldraw[fill=black, draw=black] (7.42, 1.18) rectangle (7.58, 1.02);
\filldraw[fill=black, draw=black] (8.02, 1.18) rectangle (8.18, 1.02);
\filldraw[fill=black, draw=black] (8.42, 1.18) rectangle (8.58, 1.02);
\filldraw[fill=black, draw=black] (9.02, 1.18) rectangle (9.18, 1.02);
\filldraw[fill=black, draw=black] (7.62, 0.98) rectangle (7.78, 0.82);
\filldraw[fill=black, draw=black] (8.22, 0.98) rectangle (8.38, 0.82);
\filldraw[fill=black, draw=black] (8.61, 0.98) rectangle (8.77, 0.82);
\filldraw[fill=black, draw=black] (9.22, 0.98) rectangle (9.38, 0.82);
\filldraw[fill=black, draw=black] (9.62, 0.98) rectangle (9.78, 0.82);
\filldraw[fill=black, draw=black] (10.22, 0.98) rectangle (10.38, 0.82);
\filldraw[fill=black, draw=black] (8.82, 0.78) rectangle (8.98, 0.62);
\filldraw[fill=black, draw=black] (9.42, 0.78) rectangle (9.58, 0.62);
\filldraw[fill=black, draw=black] (9.82, 0.78) rectangle (9.98, 0.62);
\filldraw[fill=black, draw=black] (10.42, 0.78) rectangle (10.58, 0.62);
\filldraw[fill=black, draw=black] (10.82, 0.78) rectangle (10.98, 0.62);
\filldraw[fill=black, draw=black] (11.42, 0.78) rectangle (11.58, 0.62);
\filldraw[fill=black, draw=black] (10.02, 0.58) rectangle (10.18, 0.41);
\filldraw[fill=black, draw=black] (10.62, 0.58) rectangle (10.78, 0.41);
\filldraw[fill=black, draw=black] (11.02, 0.58) rectangle (11.18, 0.41);
\filldraw[fill=black, draw=black] (11.62, 0.58) rectangle (11.78, 0.41);
\filldraw[fill=black, draw=black] (11.22, 0.37) rectangle (11.38, 0.21);
\filldraw[fill=black, draw=black] (11.82, 0.37) rectangle (11.98, 0.21);
\end{tikzpicture} 

 \caption{The access pattern to the eviction set. Addresses 0 to $n$ map to the same cache set.}
 \label{fig:evict-pattern}
\end{figure}        
        
\PrimeProbe is a cache attack technique that has first been used by Osvik~\etal\cite{Osvik2006}.
In a \PrimeProbe attack, the attacker constantly primes (\ie evicts) a cache set and measures how long this step took.
The amount of time the prime step took is correlated to the number of cache ways in this cache set that have been replaced by other programs.
This allows deriving whether or not a victim application performed a specific secret-dependent memory access.
Recent work has shown that this technique can even be used across virtual machine boundaries~\cite{Ristenpart2009,Zhang2011,Liu2015,Irazoqui2015SA,Maurice2017Hello}.

To prime (\ie evict) a cache set, the attacker needs $n$ addresses that map to the same cache set (\ie an \textit{eviction set}), where $n$ depends on the cache replacement policy and the number of ways of the last-level cache.
On Intel CPUs before Ivy Bridge, the cache replacement policy was Least-Recently Used (LRU), and thus it was sufficient to access $n$ addresses for an $n$-way cache. However, on newer microarchitectures, the exact cache replacement policy is unknown.
To minimize the amount of time the prime step takes, it is necessary to find a minimal $n$ combined with a fast access pattern (\ie an \textit{eviction strategy}).
Gruss~\etal\cite{Gruss2016Row} experimentally found efficient eviction strategies with high eviction rates and a small number of addresses.
We use their eviction strategy on our Skylake test machine throughout the paper. Figure~\ref{fig:evict-pattern} shows the eviction set access pattern of this eviction strategy.

A more powerful cache attack technique is \FlushReload by Yarom and Falkner~\cite{Yarom2014}. For a \FlushReload attack, attacker and victim need to share memory (\ie a shared library or page deduplication).
The attacker flushes a shared memory line from the cache to then measure the amount of time it takes to reload the cache line.
This reveals whether or not another program reloaded this exact cache line. Although \FlushReload attacks have been studied extensively~\cite{Irazoqui2015SA,Gruss2015Template,Irazoqui2015Neighbor,Guelmezoglu2015,Lipp2016,Benger2014,Inci2016,Apecechea2014a,Irazoqui2016Cross}
they are now considered impractical in the cloud as most cloud providers disabled page deduplication and thus disabled the only way to obtain shared memory in the cloud.

Pessl~\etal\cite{Pessl2016} found another attack vector that can yield an accuracy close to a \FlushReload attack without requiring shared memory. They attack the DRAM modules that are shared by all virtual machines running on the same host system.
Each DRAM module has a row buffer that holds the most recently accessed DRAM row. While accesses to this buffer are fast, accesses to other memory locations in DRAM are much slower. This timing difference can be exploited to obtain fine-grained information across virtual machine boundaries.

\subsection{Side-Channel Attacks on SGX}
There have been speculations that SGX could be vulnerable to cache side-channel attacks~\cite{Costan2016}.
In fact, Intel does not consider side channels as part of the SGX threat model and thus states that SGX does not provide any specific mechanisms to protect against side-channel attacks~\cite{sgxdeveloperref}.
However, they also explicitly state that SGX features still impair side-channel attacks.
Intel recommends using SGX enclaves to protect password managers and cryptographic keys against side channels and advertises this as a feature of SGX~\cite{Intel_HardeningSGX}.
Indeed, SGX does not provide special protection against microarchitectural attacks, its focus lies on new attack vectors arising from an untrusted operating system.
Xu~\etal\cite{Xu2015controlled} show that SGX is vulnerable to controlled channel attacks in which a malicious operating system triggers and monitors enclave page faults~\cite{sgxtutorial}.
Both attacks rely on a malicious or compromised operating system to break into an enclave.

SGX enclaves generally do not share memory with other enclaves, the operating system or other processes. Thus, \FlushReload attacks on SGX enclaves are not possible. Also, DRAM-based attacks cannot be performed from a malicious operating system, as the hardware prevents any operating system accesses to DRAM rows in the EPC. However, enclaves can mount DRAM-based attacks on other enclaves because all enclaves are located in the same physical EPC.

\subsection{Side-Channel Attacks on RSA}\label{sec:rsa-sc}

\begin{algorithm}[t]
	\SetKw{KwStep}{step}
	\SetKw{KwDownto}{downto}
	\SetKwInOut{Input}{input}\SetKwInOut{Output}{output}
	\Input{base $\mathit{b}$, exponent $\mathit{e}$, modulus $\mathit{n}$}
	\Output{$\mathit{b^{e} \mod n}$}
	\BlankLine
	$\mathit{X} \leftarrow 1$\;
	\For{$\mathit{i} \leftarrow bitlen(\mathit{e})$ \KwDownto $0$}
	{
	  $\mathit{X} \leftarrow multiply(\mathit{X}, \mathit{X})$\;
	  \If{$\mathit{e_i} = 1$}{
	    $\mathit{X} \leftarrow multiply(\mathit{X}, \mathit{b})$\;
	  }
	}
	\Return $\mathit{X}$\;
	\caption{Square-and-multiply exponentiation}\label{alg:rsa}
\end{algorithm}

RSA is widely used to create asymmetric signatures, and is implemented by virtually every TLS library, such as OpenSSL or \mbedTLS, formerly known as PolarSSL.
\mbedTLS is used in many well-known open source projects such as cURL and OpenVPN.
The small size of \mbedTLS is well suitable for the size-constrained enclaves of Intel SGX.

RSA essentially involves modular exponentiation with a private key, where the exponentiation is typically implemented as square-and-multiply, as outlined in Algorithm~\ref{alg:rsa}. 
The algorithm sequentially scans over all exponent bits. Squaring is done in each step while multiplication is only carried out if the corresponding exponent bit is set.
An unprotected implementation of square-and-multiply is vulnerable to a variety of side-channel attacks, in which an attacker learns the exponent by distinguishing the square step from the multiplication step~\cite{Yarom2014,Ge2016}.

\mbedTLS uses a windowed square-and-multiply routine for the exponentiation.
To minimize the memory footprint of the library, the official knowledge base suggests setting the window size to 1~\cite{MbedTLS_KB}.
With a fixed upper enclave memory limit in current microarchitectures, it is reasonable to follow this recommendation.
However, a window size of 1 is equivalent to the basic square-and-multiply exponentiation, as shown in Algorithm~\ref{alg:rsa}.
Liu~\etal\cite{Liu2015} showed that if an attack on a window size of 1 is possible, the attack can be extended to arbitrary window sizes.

Earlier versions of \mbedTLS were vulnerable to a timing side-channel attack on RSA-CRT~\cite{Arnaud2013}.
Due to this attack, current versions of \mbedTLS implement a constant-time Montgomery multiplication for RSA.
Additionally, instead of using a dedicated square routine, the square operation is carried out using the multiplication routine as illustrated in Algorithm~\ref{alg:rsa}. 
Thus, there is no leakage from a different square and multiplication routine as exploited in previous attacks on square-and-multiply algorithms~\cite{Aciiccmez2008,Zhang2012,Yarom2014,Liu2015}.
However, Liu~\etal\cite{Liu2015} showed that the secret-dependent accesses to the buffer $\mathit{b}$ still leak the exponent.

Boneh~\etal\cite{Boneh1998} and Blömer~\etal\cite{Blomer2003} showed that it is feasible to recover the full RSA private key if only some of either the most significant or least significant bits are known. 
Halderman~\etal\cite{Halderman2009} showed that it is even possible to recover a full RSA key if up to \SI{12}{\percent} of random bits are corrupted.  
Heninger~\etal\cite{Heninger2009} improved these results and recovered a full key for random unidirectional corruptions of up to \SI{46}{\percent}. 

%
\section{Threat Model and Attack Setup} \label{sec:threat}
%
In this section, we present our threat model.
We demonstrate a malware that circumvents SGX's and Docker's isolation guarantees. 
We successfully mount a \PrimeProbe attack on an RSA signature computation running inside a different enclave, on the outside world, and across container boundaries.

\subsection{High-Level View of the Attack}

In our threat model, both the attacker and the victim are running on the same physical machine.
The machine can either be a user's local computer or a host in the cloud.  
In the cloud scenario, the victim has its enclave running in a Docker container to provide services to other applications running on the host.
Docker containers are well supported on many cloud providers, \eg Amazon~\cite{Docker_AWS} or Microsoft Azure~\cite{Microsoft_Docker}. 
As these containers are more lightweight than virtual machines, a host can run up to several hundred containers simultaneously. 
Thus, the attacker has good chances to get a co-located container on a cloud provider.

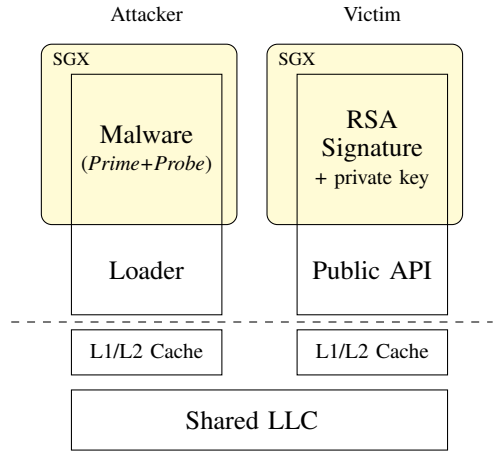
\begin{figure}
\centering
\begin{tikzpicture}[scale=0.8]
\draw (0.5,2.5) [rounded corners,fill=yellow!20] rectangle (3.75,5.5);
\draw (1,1) rectangle (3.5,5);
\node at (1,5.25) {\scriptsize SGX};
\node [text centered] at (2.25, 4) {Malware};
\node [text centered] at (2.25, 3.5) {\footnotesize (\PrimeProbe)};
\node [text centered] at (2.25, 1.75) {Loader};
\node [text centered] at (2.25, 6) {\footnotesize Attacker};
\node [text centered] at (2.25, 0.4) {\footnotesize L1/L2 Cache};
\draw (1,0) rectangle (3.5,0.75);

\begin{scope}[shift={(3.75,0)}]
\draw (0.5,2.5) [rounded corners,fill=yellow!20] rectangle (3.75,5.5);
\draw (1,1) rectangle (3.5,5);
\node at (1,5.25) {\scriptsize SGX};
\node [text centered] at (2.25, 4.25) {RSA};
\node [text centered] at (2.25, 3.75) {Signature};
\node [text centered] at (2.25, 3.25) {\footnotesize + private key};
\node [text centered] at (2.25, 1.75) {Public API};
\node [text centered] at (2.25, 6) {\footnotesize Victim};
\draw (1,0) rectangle (3.5,0.75);
\node [text centered] at (2.25, 0.4) {\footnotesize L1/L2 Cache};
\end{scope}

\draw [dashed] (0, 0.88) -- (8,0.88);
\draw (1,-1.25) rectangle (7.25,-0.25);
\node [text centered] at (4, -0.75) {Shared LLC};
\end{tikzpicture}
 \caption{The threat model: both attacker and victim run on the same physical machine in different SGX enclaves.}
 \label{fig:threat-model}
\end{figure}

Figure~\ref{fig:threat-model} gives an overview of our native setup.
The victim runs a cryptographic computation inside the enclave to protect it against any attacks.
The attacker tries to stealthily extract secrets from this victim enclave.
Both the attacker and the victim use Intel's SGX feature and are therefore subdivided into two parts, the enclave and loader, \ie the main program that instantiates the enclave.

The attack is a multi-step process that can be divided into an online and an offline phase. 
Section~\ref{sec:pp} describes the online phase, in which the attacker first locates the victim's cache sets that contain the secret-dependent data of the RSA private key. 
The attacker then monitors the identified cache sets while triggering a signature computation.
Section~\ref{sec:keyextraction} gives a detailed explanation of the offline phase in which the attacker recovers a private key from collected traces.

\subsection{Victim}\label{sec:victim}

The victim is an unprivileged program that uses SGX to protect an RSA signing application from both software and hardware attackers.
Both the RSA implementation and the private key reside inside the enclave, as suggested by Intel~\cite{Intel_HardeningSGX}.
Thus, they can never be accessed by system software or malware on the same host.
Moreover, information leakage from the enclave should not be possible due to hardware isolation and memory encryption. 
The victim uses the RSA implementation of the widely deployed \mbedTLS library, that relies on constant-time Montgomery multiplications. 
The victim application provides an API to compute a signature for provided data.

\subsection{Attacker}
The attacker runs an unprivileged program on the same host machine as the victim.
The goal of the attacker is to stealthily extract the private key from the victim enclave. 
Therefore, the attacker uses the API provided by the victim to trigger signature computations. 

The attacker targets the exponentiation step of the RSA implementation.
To perform the exponentiation in RSA, \mbedTLS uses a windowed square-and-multiply algorithm in Montgomery domain. 
The window size is fixed to 1, as suggested by the official knowledge base~\cite{MbedTLS_KB}. 
If successful, the attack can be extended to arbitrary window sizes~\cite{Liu2015}. 

To prevent information leakage from function calls, \mbedTLS uses the same function (\texttt{mpi\_montmul}) for both the square and the multiply operation (see Algorithm~\ref{alg:rsa}). 
The \texttt{mpi\_montmul} takes two parameters that are multiplied together. 
For the square operation, the function is called with the current buffer as both arguments.
For the multiply operation, the current buffer is multiplied with a buffer holding the multiplier. 
This buffer is allocated in the calling function \texttt{mbedtls\_mpi\_exp\_mod} using \texttt{calloc}.
Due to the deterministic behavior of the tlibc's \texttt{calloc} implementation, the used buffers always have the same virtual and physical addresses. Thus the buffers are always in the same cache sets.
The attacker can therefore mount a \PrimeProbe attack on the cache sets containing the buffer. 

In order to remain stealthy, all parts of the malware that contain attack code reside inside an SGX enclave.
The enclave can protect the encrypted real attack code by only decrypting it after a successful remote attestation after which the enclave receives the decryption key. 
As pages in SGX can be mapped as writable and executable, self-modifying code is possible and therefore code can be encrypted. 
Consequently, the attack is completely stealthy and invisible from anti-virus software and even from monitoring software running in ring $0$.
Note that our proof-of-concept implementation does not encrypt the attack code as this has no impact on the attack. 

The loader does not contain any suspicious code or data, it is only required to start the enclave. 
The exfiltrated data from inside the malicious enclave will only be handed to the loader in an encrypted form.
The loader may also provide a TLS endpoint through which the enclave can send encrypted data to an attacker's server.

\subsection{Operating System and Hardware}
Previous work was mostly focused on attacks on enclaves from untrusted cloud operating systems~\cite{Li2014,Baumann2015,Schuster2015,Costan2016,Aumasson2016}.
However, in our attack we do not make any assumptions on the underlying operating system, \ie we do not rely on a malicious operating system. 
Both the attacker and the victim are unprivileged user space applications.
Our attack works on a fully-patched recent operating system with no known software vulnerabilities, \ie the attacker cannot elevate its privileges.

Our only assumption on the hardware is that attacker and victim run on the same host system.
This is the case on both personal computers as well as on co-located Docker instances in the cloud. 
As SGX is currently only available on Intel's Skylake microarchitecture, it is valid to assume that the host
is a Skylake system. Consequently, we know that the last-level cache is shared between all CPU cores.

\subsection{Malware Detection}
We expect the cloud provider to run state-of-the-art malware detection software. 
We assume that malware detection software is able to monitor the behavior of containers or even inspect the content of containers. 
Moreover, the user can run anti-virus software and monitor programs inside the container.
This software can either protect the data from infections or the infrastructure from attacks. 

Standard malware detection methods are either signature-based, behavioral-based or heuristics-based~\cite{Bazrafshan2013}. 
Signature-based detection is used by virus scanners to match byte sequence insides executables against a list of such sequences extracted from known malware. 
This method is fast and rarely causes false-positives, but can only detect known malware~\cite{Sukwong2011}.   
In addition to signature-based detection, modern virus scanners implement behavior-based analysis. 
Behavior-based analysis has the potential to detect new malware by monitoring system activity, API calls, and user interactions~\cite{Sukwong2011}.

We also assume the presence of detection mechanisms using performance counters, to detect malware~\cite{Demme2013} and microarchitectural attacks~\cite{Herath2015}, which are more targeted to our attack.

%
\section{Extracting Private Key Information} \label{sec:pp}
%
In this section, we describe the online phase of our attack. 
We first build primitives necessary to mount this attack. 
Then we show in two steps how to locate and monitor cache sets to extract private key information. 

\subsection{Attack Primitives in SGX}
Successful \PrimeProbe attacks require two primitives: a high-resolution timer to distinguish cache hits and misses and a method to generate an eviction set for an arbitrary cache set.
Due to the restrictions of SGX enclaves, we cannot rely on existing \PrimeProbe implementations, and therefore we require new techniques to build a malware from within an enclave. 

        \subsubsection{High-resolution Timer}
The unprivileged \texttt{rdtsc} and \texttt{rdtscp} instructions, which read the timestamp counter, are usually used for fine-grained timing outside enclaves.
In \SGXone, these instructions are not permitted inside an SGX enclave, as they might cause a VM exit~\cite{Intel_vol3}. 
Therefore, we have to rely on a different timing source.

Lipp~\etal\cite{Lipp2016} demonstrated a counting thread as a high-resolution alternative on ARM where no unprivileged high-resolution timer is available. 
The idea is to have a dedicated thread incrementing a global variable in an endless loop. 
As the attacks only rely on accurate timing differences and not on absolute timestamps, this global variable serves directly as the timing source.

We require a minimum resolution in the order of \SI{10}{cycles} to reliably distinguish cache hits from misses as well as DRAM row hits from row conflicts.
To achieve the highest number of increments, we handcraft the counter increment in inline assembly.
According to Intel~\cite{Intel_opt}, the fastest instructions on the Skylake microarchitecture are \texttt{inc} and \texttt{add} with both a latency of \SI{1}{cycle} and a throughput of \SI{0.25}{cycles/instruction} when executed with a register as an operand. 
The counter variable has to be accessible across threads, thus it is necessary to store the counter variable in memory. 
Memory addresses as operands incur an additional cost of approximately \SI{4}{cycles} due to L1 cache access times~\cite{Intel_opt}. 
To reduce the cost of the \texttt{jmp} instruction, we tried to unroll the loop up to the point where we get the most increments per CPU cycle.
However, our experiments showed that the unrolling tends to rather have negative effects on the performance.  
On our test machine, the code from Listing~\ref{fig:counting-thread} achieves one increment every \SI{4.7}{cycles}, which is an improvement of approximately \SI{2}{\percent} over the assembly code generated by \texttt{gcc} on the highest optimization level (\texttt{-O3}).

\begin{listing}[t]
\begin{lstlisting}[language={[x64]Assembler}]
mov &counter, %rcx
1: incl (%rcx)
jmp 1b
\end{lstlisting}
\caption{A counting thread that emulates \texttt{rdtsc}.}
\label{fig:counting-thread}
\end{listing}

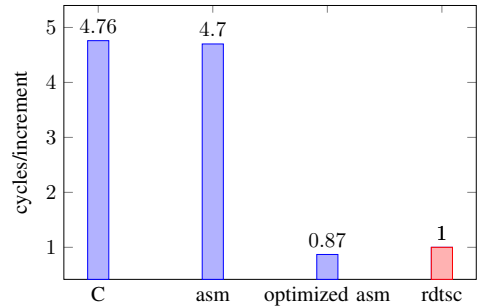
\begin{figure}
\centering
\begin{tikzpicture}[scale=0.8]
  \begin{axis}[
	    legend style={at={(0.5,-0.45)},
	    anchor=north,legend columns=-1},
            xticklabels={C, asm, optimized asm,rdtsc},
            xticklabel style={text height=1.5ex},
	    ylabel={cycles/increment},
            xtick={1,2,3,4},
            yscale=0.8,
            ymax=5.4,
	    nodes near coords,
	    nodes near coords align={anchor=south}
          ]
            \addplot[ybar,fill=lblue,draw=blue] coordinates {
                (1,   4.76)
                (2,  4.7)
                (3,  0.87)
                (4,  1)
                };
	    \addplot[ybar,fill=lred,draw=red] coordinates {
                (4,  1)
            };
  
  \end{axis}
\end{tikzpicture}
\caption{Comparisons of different variants for the counter thread to Intel's native timestamp counter as the baseline.}
\label{fig:time-perf}
\end{figure}

We can improve the performance---and thus the resolution---further, by exploiting the fact that only the counting thread is writing to the counter variable. 
Reading the counter variable from memory is therefore never necessary as the value will not be changed by any other thread. 
To gain a higher performance from this observation, we have to eliminate the CPU's read access to the counter variable. 
Executing arithmetic operations directly on the memory location is thus not an option anymore, and it is necessary to perform any operation with data dependency on a CPU register.
Therefore, we introduce a ``shadow counter variable'' which is always held in a CPU register. 
The arithmetic operation (either \texttt{add} or \texttt{inc}) is performed using this register as the operand, unleashing the low latency and throughput of these instructions.
As registers cannot be shared across threads, the shadow counter has to be moved to memory using the \texttt{mov} instruction after each increment. 
Similar to the \texttt{inc} and \texttt{add} instruction, the \texttt{mov} instruction has a latency of \SI{1}{cycle} and a throughput of \SI{0.5}{cycles/instruction} when copying a register to a memory location.
Listing~\ref{fig:counting-thread2} shows the improved counting thread. 
This counting thread is capable of incrementing the variable by one every \SI{0.87}{cycles}, which is an improvement of \SI{440}{\percent} over the code in Listing~\ref{fig:counting-thread}.
In fact, this version is even \SI{15}{\percent} faster than the native timestamp counter, thus giving us a reliable timing source that even has a higher resolution.
This new method might open new possibilities of side-channel attacks that leak information through timing on a sub-\texttt{rdtsc} level. 
Figure~\ref{fig:time-perf} shows the performance comparison of the C version, the assembly version, the optimized assembly version, and the native timestamp counter as a baseline. 
Although the method with the shadow counter has the most instructions in the loop body, and an increase of \SI{100}{\percent} in code size compared to Listing~\ref{fig:counting-thread}, it has the best performance. 
Due to multiple execution units, pipelining, and the absence of memory dependencies, one increment can be carried out in less than \SI{1}{cycle} on the Skylake microarchitecture even though each instruction has a latency of \SI{1}{cycle}~\cite{Fog2016}.

\begin{listing}[t]
\begin{lstlisting}[language={[x64]Assembler}]
mov &counter, %rcx
1: inc %rax
mov %rax, (%rcx)
jmp 1b
\end{lstlisting}%
\caption{The improved fast counting thread that acts as the emulation of \texttt{rdtsc}.}%
\label{fig:counting-thread2}
\end{listing}

        \subsubsection{Eviction Set Generation}
\PrimeProbe relies on eviction sets, \ie we need to find virtual addresses that map to the same cache set.
An unprivileged process cannot translate virtual to physical addresses and therefore cannot simply search for virtual addresses that fall into the same cache set. 
This limitation also applies to enclaves, as they are always unprivileged. 
Liu~\etal\cite{Liu2015} and Maurice~\etal\cite{Maurice2017Hello} demonstrated algorithms to build eviction sets using large pages by exploiting the fact that the virtual address and the physical address have the same lowest 21 bits.
At least in the current version, SGX does not support large pages, making this approach inapplicable. 
Oren~\etal\cite{Oren2015} and Gruss~\etal\cite{Gruss2016Row} demonstrated fully automated methods to generate an eviction sets for a given virtual address. 
However, the method of Oren~\etal\cite{Oren2015} uses a slow pointer-chasing approach and needs to find an eviction set without any assumptions, consuming more time.
The method by Gruss~\etal\cite{Gruss2016Row} has the overhead of finding an eviction strategy and eviction set without any assumptions. Thus, applying their approach for our purposes would consume multiple hours on average before even starting the actual \PrimeProbe attack.

We propose a new method to recover the cache set from a virtual address without relying on large pages. 
The method requires that an array within an SGX enclave is backed by physically contiguous pages. 
We verified that we have contiguous pages by inspecting Intel's SGX driver for Linux~\cite{Intel2016intelsgxlinux}. 
When initializing a new enclave, the function \texttt{isgx\_page\_cache\_init} creates a list of available physical pages for the enclave. 
These pages start at a base physical address and are contiguous.  
If a physical page is mapped, \eg due to a page fault, the function \texttt{isgx\_alloc\_epc\_page\_fast} removes and returns the head of the list.

The idea is to exploit the DRAM timing differences that are due to the DRAM organization and to use the DRAM mapping functions~\cite{Pessl2016} to recover physical address bits.
Alternately accessing two virtual addresses that map to the same DRAM bank but a different row is significantly slower than any other combination of virtual addresses.
For the first address of a DRAM row, the least-significant \SI{18}{bits} of the physical address are \zero, because the row index only uses physical address bits $18$ and upwards.
Thus, we scan memory sequentially for an address pair in physical proximity that causes a \textit{row conflict}. 
As SGX enclave memory is allocated in a contiguous way we can perform this scan on virtual addresses.

A virtual address pair that causes row conflicts at the beginning of a row satisfies the following constraints:
\begin{compactenum}
 \item The bank address (BA), bank group (BG), rank, and channel must be the same for both virtual addresses. Otherwise, a row conflict is not possible.
 \item The row index must be different for both addresses.
 \item The difference of the two physical addresses (of the virtual addresses) has to be at least \SI{64}{B} (the size of one cache line) but should not exceed \SI{4}{kB} (the size of one page).
 \item Physical address bits $6$ to $22$ have the same known value, all 0 for the higher address and all 1 for the lower address, as only bits in this range are used by the mapping function.
\end{compactenum}
For all virtual addresses satisfying these constraints, bits $6$ to $22$ have a known value. Thus, we know the exact cache set for these virtual addresses.

Table~\ref{fig:dram-mapping} shows the reverse-engineered DRAM mapping function for our test machine, an Intel Core i5-6200U with \SI{12}{GB} main memory. 
The row index is determined by the physical address bits starting from bit $18$. 
\begin{table}[t]
\centering
\caption{Reverse-engineered DRAM mapping functions using the method from Pessl~\etal\cite{Pessl2016}.}
\label{fig:dram-mapping}
\bgroup
\setlength{\tabcolsep}{0.1em}
\scriptsize
\begin{tabular}{|l|c||c|c|c|c|c|c|c|c|c|c|c|c|c|c|c|c|c|}
  \cline{3-19}
\multicolumn{2}{c|}{} & \multicolumn{17}{c|}{Address Bit}\\
 \cline{3-19} \multicolumn{2}{c|}{} & 2 & 2 & 2 & 1 & 1 & 1 & 1 & 1 & 1 & 1 & 1 & 1 & 1 & 0 & 0 & 0 & 0 \\ 
         \multicolumn{2}{c|}{} & 2 & 1 & 0 & 9 & 8 & 7 & 6 & 5 & 4 & 3 & 2 & 1 & 0 & 9 & 8 & 7 & 6 \\ 
\hline \hline \multirow{6}{*}{2 DIMMs}   & Channel &        &        &        & $\xor$ & $\xor$ &        &        &        &        & $\xor$ & $\xor$ & \phantom{$\xor$} & \phantom{$\xor$} & $\xor$ & $\xor$ &        &  \phantom{$\xor$} \\  
\cline{2-19}                                                                                                                                
                                            & BG0 &        &        &        &        &        &        &        &        & $\xor$ &         &        &        &        &        &        & $\xor$ & \\
\cline{2-19}                                                                                                                                                   
                                            & BG1 & $\xor$ &        &        &        & $\xor$ &        &        &        &        &         &        &        &        &        &        &        & \\
\cline{2-19}                                                                                                                                                                              
                                            & BA0 &        &        &        & $\xor$ &        &        &        & $\xor$ &        &         &        &        &        &        &        &        & \\
\cline{2-19}                                                                                                                                                                              
                                            & BA1 &        & $\xor$ &        &        &        & $\xor$ &        &        &        &         &        &        &        &        &        &        & \\
\cline{2-19}                                                                                                                                                                              
                                           & Rank &        &        & $\xor$ &        &        &        & $\xor$ &        &        &         &        &        &        &        &        &        & \\
                                            
\hline
\end{tabular}
\vspace{0.5em}
\egroup
\end{table}

To find address pairs fulfilling the aforementioned constraints, we modeled the mapping function and the constraints as an SMT problem and used the Z3 theorem prover~\cite{DeMoura2008} to provide models satisfying the constraints. 
The model we found yields pairs of physical addresses where the upper address is \SI{64}{B} apart from the lower one. 
There are four such address pairs within every \SI{4}{MB} block of physical memory such that each pair maps to the same bank but a different row.
The least-significant bits of the physical address pairs are either (\texttt{0x3fffc0}, \texttt{0x400000}), (\texttt{0x7fffc0}, \texttt{0x800000}), (\texttt{0xbfffc0}, \texttt{0xc00000}) or (\texttt{0xffffc0}, \texttt{0x1000000}) for the lower and higher address respectively. 
Thus, at least \SI{22}{bits} of the higher addresses least-significant bits are 0. 

\begin{figure}[t]
        \centering
        \begin{tikzpicture}[scale=0.85]
        \begin{axis}[
        xlabel={Array index [kB]},
        ylabel={Access time [CPU cycles]}]
        \addplot+ table [x expr=\coordindex+1,y index=0, mark=none, smooth] {data/dram-time.csv};
        \addplot table [x index=0,y index=1,only marks,draw=none] {data/dram-time-highlight.csv};
        \end{axis}
        \end{tikzpicture}
 \caption{Access times when alternately accessing two addresses which are \SI{64}{B} apart. The (marked) high access times indicate row conflicts.}
 \label{fig:dram-timing}
\end{figure}
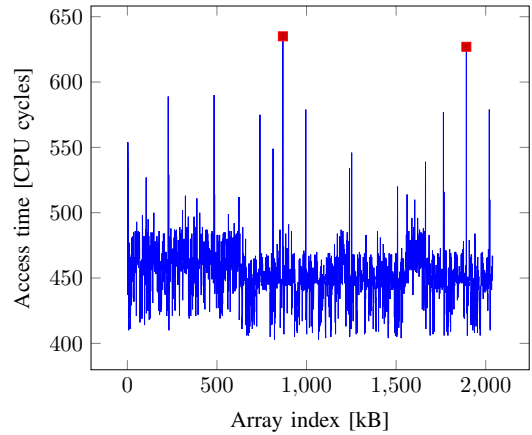

Figure~\ref{fig:dram-timing} shows the average access time for address pairs when iterating over a \SI{2}{MB} array. 
The highest two peaks show row conflicts, \ie the row index changes while the bank, rank, and channel stay the same. 
As the cache set is determined by the bits $6$ to $17$, the higher address has the cache set index 0 at these peaks.   
Based on the assumption of contiguous memory, we can generate addresses mapping to the same cache set by adding multiples of \SI{256}{KB} to the higher address.

As the last-level cache is divided into multiple parts called cache slices, there is one cache set per slice for each cache set index.
Thus, we will inherently add addresses to our generated eviction set that have no influence on the eviction although they have the correct cache set index. 
For the eviction set, it is necessary to only use addresses that map to the same cache slice. 
However, to calculate the cache slice from a physical address, all bits of the physical address are required~\cite{Maurice2015RAID}.

As we are not able to directly calculate the cache slice, we use another approach. 
We add our calculated addresses from the correct cache set to our eviction set until the eviction rate is sufficiently high.
Then, we try to remove single addresses from the eviction set as long as the eviction rate does not drop. 
Thus, we remove all addresses that do not contribute to the eviction, and the result is a minimal eviction set.
Algorithm~\ref{alg:eviction-set} shows the full algorithm to generate a minimal eviction set. 
Our approach takes on average 2 seconds per cache set, as we already know that our addresses map to the correct cache set.
This is nearly three orders of magnitude faster than the approach of Gruss~\etal\cite{Gruss2016Row}.

\begin{algorithm}[t]
	\SetKw{KwStep}{step}
	\SetKwInOut{Input}{input}\SetKwInOut{Output}{output}
	\Input{$\mathit{memory}$: char[$8 \times 1024 \times 1024$], 
		$\mathit{set}$: int}
	\Output{$\mathit{eviction\_set}$: char*[$\mathit{n}$]}
	\BlankLine
	$\mathit{border} \leftarrow 0$\;
	$\mathit{border\_index} \leftarrow 0$\;
	\For{$i \leftarrow \mathit{0xFC0}$ \KwTo $4 \times 1024 \times 1024$ \KwStep $4096$}
	{
	  $\mathit{time} \leftarrow$ hammer($\mathit{memory[i]}$, $\mathit{memory[i + 64]}$)\;
	  \If{$\mathit{time} > \mathit{border}$}{
	  $\mathit{border} \leftarrow \mathit{time}$\;
	  $\mathit{border\_index} \leftarrow i + 64$\;
	  }
	}
	$addr \leftarrow (\&\mathit{memory}[\mathit{border\_index}]) + \mathit{set} \ll 6$\;
	$n \leftarrow 0$\;
	\Repeat{$\mathit{eviction} > 99\%$}{
	  $\mathit{full\_set}[n] \leftarrow addr + n \times \SI{256}{KB}$\;
	  $\mathit{eviction} \leftarrow$ evict($\mathit{full\_set}$, $n$)\;
	  $n \leftarrow n + 1$\;
	}
	\For{$i \leftarrow 0$ \KwTo $n$}{
	  $\mathit{removed} \leftarrow \mathit{full\_set}[i]$\;
	  $\mathit{full\_set}[i] \leftarrow \mathit{NULL}$\;
	  \lIf{evict($\mathit{full\_set}$, $n$) $ < 99\%$}{$\mathit{full\_set}[i] \leftarrow \mathit{removed}$}
	}
	$len \leftarrow 0$\;
	\For{$i \leftarrow 0$ \KwTo $n$}{
	  \If{$\mathit{full\_set}[i] \neq \mathit{NULL}$}{
	    $\mathit{eviction\_set}[len] \leftarrow \mathit{full\_set}[i]$\;
	    $\mathit{len} \leftarrow \mathit{len} + 1$\;
	  }
	}
	\caption{Generating the eviction set}\label{alg:eviction-set}
\end{algorithm}

     \subsection{Identifying Vulnerable Sets}
Now that we have a reliable high-resolution timer and a method to generate eviction sets, we can mount the first stage of the attack and identify the vulnerable cache sets. 
As we do not have any information on the virtual or physical addresses of the victim, we have to scan the last-level cache for characteristic patterns that correspond to the signature process. 
We consecutively mount a \PrimeProbe attack on every cache set while the victim is executing the exponentiation step. 
This allows us to log cache misses due to a victim's activity inside the monitored cache set. 

\begin{figure}[t]
	\centering
	\begin{tikzpicture}[scale=0.85]
	\pgfplotsset{every axis legend/.append style={at={(0.5,1.2)},anchor=north,draw=none}}
	\begin{axis}[
	legend columns=2,
	xlabel={Access time [CPU cycles]},
	ylabel={Number of cases},
	height=5.5cm,
	width=10cm,
	ybar,
	ymin=1,
	xmin=0,
	xmax=800,
	ybar interval,
	xmajorgrids=false,
	legend image code/.code={\draw[#1] (0cm,-0.2cm) rectangle (0.6cm,0.2cm);},
	xtick={0,100,...,800},
	xticklabels={0,100,200,300,400,500,600,700,800},
	]
	\addplot+[blue,fill=lblue,area legend] table[x=cycles,y=cached] {data/probe.csv};
	\addlegendentry{no cache activity \quad \, }
	\addplot+[red,fill=lred,area legend] table[x=cycles,y=evicted] {data/probe.csv};
	\addlegendentry{cache activity}
	\end{axis}
	\end{tikzpicture}

 \caption{Histogram showing the runtime of the prime step for cache activity in the same set and no cache activity in the same set.}
 \label{fig:pptime}
\end{figure}
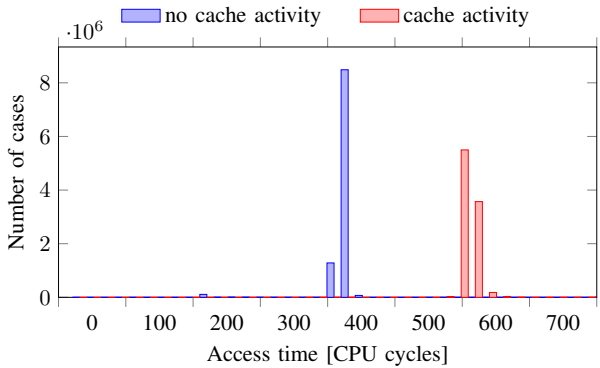

First, we fill the cache lines of this cache set with the eviction set using the access pattern shown in Algorithm~\ref{alg:primeprobe}. 
This step is called the \textit{prime} step. 
We expect our addresses to stay in the cache if the victim has no activity in this specific cache set.
Second, we measure the runtime of this algorithm to infer information about the victim. 
We refer to this step as the \textit{probe} step. 
Figure~\ref{fig:pptime} shows the timings of a probe step with and without cache activity of the victim. 
If there was no activity of the victim inside the cache set, the probe step is fast as all addresses of the eviction set are still cached. 
If we encounter a high timing, we know that there was activity inside the cache set and at least one of our addresses was evicted from the cache set. 
For all following measurements, the probe step also acts as the prime step. 
The measurement ensures that the eviction set is cached again for the next round.

We can identify multiple cache sets showing this distinctive pattern which consists of three parts. 
The start of an exponentiation is characterized by a high usage of the cache set due to clearing and initialization of the used buffers. 
It is followed by the actual exponentiation that depends on the secret exponent.
The exponentiation ends with another high peak where the buffer is cleared, followed by no cache misses anymore, \ie it is only influenced by background noise. 

To automatically find these sets, we apply a simple peak detection to find the rightmost peak. 
If we can identify another peak before that within a certain range, we assume that this cache set is used by our target buffer.
Depending on the size of the RSA exponent, we get multiple cache sets matching this pattern. 
Our experiments showed that using identified sets which are neither at the beginning nor at the end yields good results in the actual attack. 
The first and last cache set might be used by neighboring buffers and they are more likely to be prefetched~\cite{Yarom2014,Gruss2015Template}. Thus, they are more prone to measurement errors.

        \subsection{Monitoring Vulnerable Sets}
Once we have identified a cache set which is used by the exponentiation, we can collect the actual traces. 
The measurement method is the same as for detecting the vulnerable cache sets, \ie we again use \PrimeProbe. 
Due to the deterministic behavior of the heap allocation, the address of the attacked buffer does not change on consecutive exponentiations. 
Thus, we can collect multiple traces of the signature process. 

To maintain a high sampling rate, we keep the post-processing during the measurements to a minimum. 
Moreover, it is important to keep the memory activity at a minimum to not introduce additional noise on the cache. 
Thus, we only save the timestamps of the cache misses for further post-processing. 

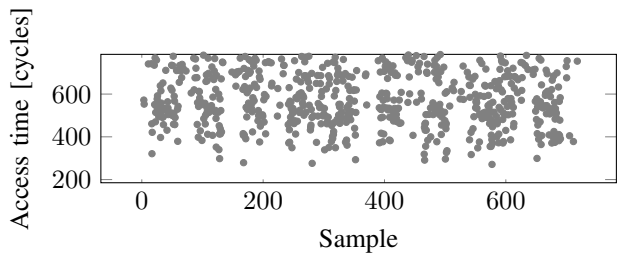
\begin{figure}
 \centering
      \begin{tikzpicture}
        \begin{axis}[
        xscale=1,
        yscale=0.3,
        ymax=700,
        enlarge y limits=0.2,
        xlabel style={yshift=-10pt},
        xlabel={Sample},
        ylabel={Access time [cycles]}
        ]
        \addplot[mark=*, mark size=1.2,gray,fill=gray,mark options={yscale=3.3}] table [x index=0,y index=1, only marks] {data/monitor.csv};
        \end{axis}
      \end{tikzpicture}
 \caption{Dense areas indicate a high cache-hit rate, white areas are intervals with cache misses.}
 \label{fig:ts}
\end{figure}

Figure~\ref{fig:ts} shows the measurement for one run of the signature algorithm. 
We can see intervals with multiple cache misses and intervals without cache misses, corresponding to high cache usage and no cache usage of the victim, respectively. 
As a cache miss takes longer than a cache hit, the effective sampling rate varies depending on the number of cache misses. 
We have to consider this effect in the post-processing as it induces a non-constant sampling interval.

\begin{algorithm}[t]
	\SetKwInOut{Input}{input}
	\Input{$\mathit{n}$: int, \\
$\mathit{addrs}$: int[$\mathit{n}$]}
	\BlankLine
	\For{$i \leftarrow 0$ \KwTo $\mathit{n}-2$}{
		*addrs[i];\\
		*addrs[i+1];\\
		*addrs[i+2];\\
		*addrs[i];\\
		*addrs[i+1];\\
		*addrs[i+2];\\
	}
	\caption{Attacker accessing a set.}\label{alg:primeprobe}
\end{algorithm}

%
\section{Recovering the Private Key}\label{sec:keyextraction}
%
In this section, we describe the offline phase of our attack: recovering the private key from the recorded traces of the victim enclave. 
This can either be done inside the malware's enclave or on the attacker's server. 

Ideally, one would combine multiple traces by aligning them and averaging out noise. The more traces are combined, the more noise is eliminated. From the resulting averaged trace one can easily extract the private key.
However, the traces obtained in our attack are affected by several noise sources.
Most of them alter the timing, making trace alignment difficult. 
Among them are interrupts which lead to context switches and therefore descheduling of the attacker or the victim.
Other sources of noise include unrelated activity on the enclave's cache sets and varying CPU clock frequency due to power management. 
Although methods exist for aligning such traces~\cite{Woudenberg2011,Muijrers2011}, we opt for a different strategy. 
Instead of attempting to align traces beforehand, we pre-process all traces individually and extract a partial key out of each trace. 
These partial keys likely suffer from random insertion and deletion errors as well as from bit flips. 
To eliminate those errors, multiple partial keys are combined in the key recovery phase. 
This approach has much lower computational overhead than trace alignment since key recovery is performed on partial keys of length \SI{4}{\KB} instead of full traces containing several thousand measurements. 

Key recovery comes in three steps. First, traces are pre-processed. Second, a partial key is extracted from each trace. Third, the partial keys are merged to recover the private key. 

        \subsection{Pre-processing}

\begin{figure*}
\centering
        \begin{tikzpicture}
        \begin{axis}[
        hide axis,
        xscale=2,
        yscale=1.2,
        ymax=2000,
        height=4.5cm,
        width=10cm
        ]
        \addplot[mark=square*, mark size=1.1, yshift=80, yscale=52, fill=black!20,opacity=0.3,draw=white] table [x index=0,y index=1, only marks] {data/trace.key1.csv};
        \addplot[mark=*, mark size=0.35, yshift=60,black!80,fill=black!80,mark options={yscale=0.83,xscale=0.5}] table [x index=0,y index=1, only marks] {data/trace.orig.csv};
        \addplot[black!85] table [x expr=\coordindex+1,y index=0, mark=none, smooth] {data/trace.sampled.csv};
        
        \addplot[
	  line width=0cm,
	  yshift=55,
	  draw=none,
	  nodes near coords, 
	  point meta=explicit symbolic, 
	  every node near coord/.style={anchor=south,font=\scriptsize} 
	  ] table [meta index=2] {data/trace.keylabel.csv};
        \end{axis}
        \end{tikzpicture}
 \caption{On top is one trace's raw measurement over \SI{4000000}{cycles}. The peaks in the resampled trace on the bottom clearly indicate \ones.}
 \label{fig:real-trace}
\end{figure*}
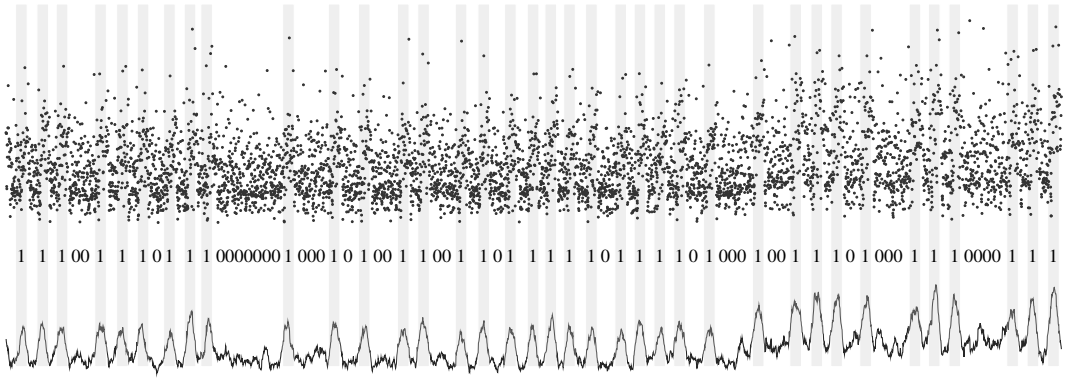

In the pre-processing step we filter and resample raw measurement data.
Figure~\ref{fig:real-trace} shows a trace segment before (top) and after pre-processing (bottom).
High values in the raw measurement data correspond to cache misses whereas low values indicate cache hits.
Timing measurements have a varying sample rate. This is because a cache miss delays the next measurement while cache hits allow more frequent measurements.
To simplify the subsequent steps, we convert the measurements to a constant sampling rate.
Therefore, we specify sampling points \SI{1000}{cycles} apart.
At each sampling point we compute the normalized sum of squared measurements within a \SI{10000}{cycle} window.
Squaring the measurements is necessary to account for the varying sampling rate.
If the measurements exceed a certain threshold, they are considered as noise and are discarded. 
If too few measurements are available within a window, \eg due to an interrupt, we apply linear interpolation. 
The resulting resampled trace shows high peaks at locations of cache misses, indicating a \one in the RSA exponent, as shown in Figure~\ref{fig:real-trace} on the bottom.

\subsection{Partial Key Extraction}
To automatically extract a partial key from a resampled trace, we first run a peak detection algorithm.
We delete duplicate peaks, \eg peaks where the corresponding RSA multiplications would overlap in time.
We also delete peaks that are below a certain adaptive threshold, as they do not correspond to actual multiplications.
Using an adaptive threshold is necessary since neither the CPU frequency nor our timing source (the counting thread) is perfectly stable.
The varying peak height is shown in the right third of Figure~\ref{fig:real-trace}.
The adaptive threshold is the median over the 10 previously detected peaks.
If a peak drops below \SI{90}{\percent} of this threshold, it is discarded.
The remaining peaks correspond to the \ones in the RSA exponent and are highlighted in Figure~\ref{fig:real-trace}.
\zeros can only be observed indirectly in our trace as square operations do not trigger cache activity on the monitored sets.
\zeros appear as time gaps in the sequence of \one peaks, thus revealing all partial key bits.
Note that since \zeros correspond to just one multiplication, they are roughly twice as fast as \ones. 

A partial key might suffer from bit flips, random insertions, and deletions, when compared to the correct key. 
When a correct peak is falsely discarded, the corresponding \one is interpreted as two \zeros.
Likewise, if noise is falsely interpreted as a \one, this cancels out two \zeros.
Moreover, if the attacker is not scheduled, we miss certain key bits in the trace. 
If the victim is not scheduled, we see a region of cache inactivity in the measurement that cannot be distinguished from true \zeros. 
Finally, if both the attacker and the victim are descheduled, this gap does not show up prominently in the trace since the counting thread is also suspended by the interrupt. 
This is in fact an advantage of a counting thread over the use of the native timestamp counter.
The remaining errors in the partial keys are corrected in the final key recovery.

        \subsection{Final Key Recovery}\label{sec:keyrecovery}
In the final key recovery, we merge multiple partial keys to obtain the full key. 
We quantify partial key errors using the edit distance~\cite{Levenshtein1966Binary}. 
The edit distance between a partial key and the correct key gives the number of bit insertions, deletions and flips necessary to transform the partial key into the correct key.

\begin{table}
\centering
\caption{Bit-wise key recovery over five partial keys. }
\label{tbl:keyrecovery}
\begin{tabular}{|c|l|} \hline
No. & Recovered key\\ \hline
1 & \underline{10111}\textcolor{blue}{\textbf 1}10001100110010111101010000100... \\ \hline
2 & \underline{10111}\textcolor{red}{\textit 0}11000111001100101101101010000... \\ \hline
3 & \underline{10111}\textcolor{blue}{\textbf 1}10001110011001011110101000010... \\ \hline
4 & \underline{10111}\textcolor{blue}{\textbf 1}10001110001100101111010100001... \\ \hline
5 & \underline{10111}\textcolor{blue}{\textbf 1}10001110011001011100010100001... \\ \hline
	\end{tabular}
\end{table}

Algorithm~\ref{alg:keyrecovery} shows the pseudo code for the final key recovery.
The full key is recovered bitwise, starting from the most-significant bit.
The correct key bit is the result of the majority vote over the corresponding bit in all partial keys.
Before proceeding to the next key bit, we correct all wrong partial keys which did not match the recovered key bit. 
To correct the current bit of the wrong partial key, we compute the edit distance to all partial keys that won the majority vote.
To reduce performance overhead, we calculate the edit distance, not over the whole partial keys but only over a lookahead window of a few bits. 
The output of the edit distance algorithm is a list of actions necessary to transform one key into the other.
We apply these actions via majority vote until the key bit of the wrong partial key matches the recovered key bit again. 
Table~\ref{tbl:keyrecovery} gives an example where the topmost 5 bits are already recovered (underlined). 
The sixth key bit is recovered as \one, since all partial key bits---except for the second one---are \one (bold). 
The incorrect \zero of the second partial key is deleted before proceeding to the next bit. 
This procedure is repeated for all key bits until the majority of partial keys reached the last bit.

\begin{algorithm}[t]
	\SetKwInOut{Input}{input}
	\SetKwInOut{Output}{output}
	\Input{$\mathit{keys}$: boolean[], $\mathit{lookahead}$: int}
	\Output{$\mathit{key}$: boolean[]}
	\BlankLine
	$\mathit{key} \leftarrow []$;\\
	$\mathit{i} \leftarrow 0$;\\
	
	\While{True}{
		$\mathit{keybit} \leftarrow $ majority($\mathit{keys}$, $\mathit{i}$);\\
		\If{$\mathit{keybit} = \perp$}{
			\Return{$\mathit{key}$};\\
		}
		$\mathit{key[i]} \leftarrow \mathit{keybit}$;\\
		$\mathit{correct} \leftarrow$ \{\};\\
		$\mathit{wrong} \leftarrow$ \{\};\\
		\ForEach{$\mathit{k}$ in $\mathit{keys}$}{
			\eIf{$\mathit{k[i]} = \mathit{keybit}$} {
				$\mathit{correct} \leftarrow \mathit{correct} \cup \mathit{k}$;\\
			}{
				$\mathit{wrong} \leftarrow \mathit{wrong} \cup \mathit{k}$;\\
			}
		}
		\ForEach{$\mathit{kw}$ in $\mathit{wrong}$}{
			$\mathit{actions} \leftarrow$ \{\};\\
			\ForEach{$\mathit{kc}$ in $\mathit{correct}$}{
				$\mathit{actions} \leftarrow \mathit{actions} \cup$ EditDistance($\mathit{kw[i:i+lookahead]}$,\\
\qquad\qquad\qquad$\mathit{kc[i:i+lookahead]}$); \\
			}
			$\mathit{ai} \leftarrow 0$;\\
			\While{$\mathit{kw[i]} \neq \mathit{keybit}$} 				{
				$\mathit{action} \leftarrow$ majority($\mathit{actions}$, $\mathit{ai}$);\\
				apply $\mathit{action}$ to $\mathit{kw[i]}$;\\
				$\mathit{ai}$++;\\
			}
			
		}
		
		$\mathit{i}$++;\\
	}

	\BlankLine
	\SetKw{KwFunction}{function}
	
	\KwFunction majority($\mathit{set}$, $\mathit{idx}$)
	\Begin{
		$\mathit{counter}[] \leftarrow 0$;\\
		\ForEach{$\mathit{array}$ in $\mathit{set}$} {
			$\mathit{element} \leftarrow \mathit{array[idx]}$;\\
			increment $\mathit{counter[element]}$;\\
			
		}
		\Return $\mathit{element}$ with max. $\mathit{counter}$;\\
	}
	\caption{RSA private key recovery.}\label{alg:keyrecovery}
\end{algorithm}        

%
\section{Evaluation} \label{sec:evaluation}
%

In this section, we evaluate the presented methods by building a malware enclave attacking a co-located enclave that acts as the victim. 
As discussed in Section~\ref{sec:victim}, we use \mbedTLS, in version 2.3.0. 
The small code and memory footprint and self-containment of \mbedTLS makes it easy to use in SGX enclaves. 

  \subsection{RSA Key Sizes and Exploitation}
For the evaluation, we attack a 4096-bit RSA key as this provides long-term security, based on the recommendation of NIST~\cite{Barker2015}.
Higher bit sizes are rarely used outside tinfoil-hat environments. 

Table~\ref{tab:keysizes} shows various RSA key sizes and the corresponding buffer sizes in \mbedTLS.
The runtime of the multiplication function increases exponentially with the size of the key. Hence, larger keys improve the measurement resolution of the attacker. In terms of cache side-channel attacks, large RSA keys do not provide higher security but degrade side-channel resistance~\cite{Walter2003,Yarom2014recovering,Yarom2014,Pereida2016}.

\begin{table}
\centering
\caption{RSA key sizes and the corresponding CPU cycles to execute one multiplication.}
\label{tab:keysizes}
\begin{tabular}{lrrr}
\toprule
 Key size & Buffer size & Cache sets & CPU cycles \\ \midrule
 \SI{1024}{b} & \SI{136}{B} & \SIx{3} & \SIx{1764} \\
 \SI{2048}{b} & \SI{264}{B} & \SIx{5} & \SIx{6624} \\
 \SI{4096}{b} & \SI{520}{B} & \SIx{9} & \SIx{25462} \\
 \SI{8192}{b} & \SI{1032}{B} & \SIx{17} & \SIx{100440} \\ \bottomrule
\end{tabular}
\end{table}

        \subsection{Native Environment}
We use a Lenovo ThinkPad T460s running Ubuntu 16.10. 
This computer supports \SGXone using Intel's SGX driver. 
The hardware details for the evaluation are shown in Table~\ref{tab:setup}. 
Both the attacker enclave and the victim enclave are running on the same machine. 
We trigger the signature process using the public API of the victim's enclave. 

\begin{table}[t]
	\centering
	\caption{Experimental setup.}\label{tab:setup}
	\begin{tabular}{llll}
		\toprule
		Environment & CPU model & Cores & LLC associativity\\
		\midrule
		Native & Core i5-6200U & 2 & 12 \\
		Docker & Core i5-6200U & 2 & 12 \\
		\bottomrule
	\end{tabular}
\end{table}

\begin{figure}[t]
 \centering
 \includegraphics[width=8cm]{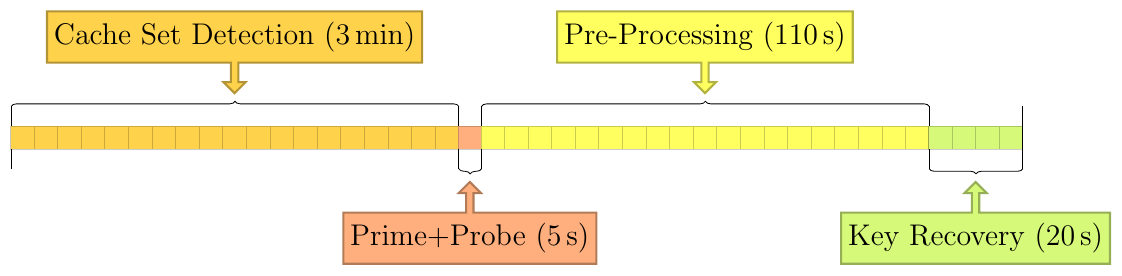}
 \caption{A high-level overview of the average times for each step of the attack.}
 \label{fig:attack-times}
\end{figure}

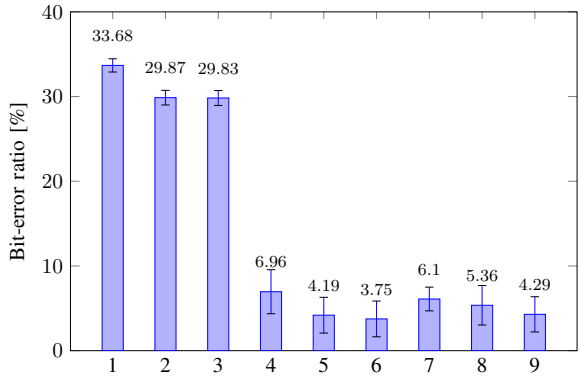
\begin{figure}[t]
\centering
\begin{tikzpicture}[scale=0.8]
  \begin{axis}[
	    legend style={at={(0.5,-0.45)},
	    anchor=north,legend columns=-1},
            xticklabels={1,2,3,4,5,6,7,8,9},
            xticklabel style={text height=1.5ex},
	    ylabel={Bit-error ratio [\%]},
            xtick={1,2,3,4,5,6,7,8,9},
            ymax=40,
            ymin=0,
            yscale=0.8,
            width=10cm,
	    nodes near coords,
	    nodes near coords align={vertical},
	    enlarge x limits=0.1,
	    every node near coord/.append style={font=\footnotesize,yshift=8pt}
          ]
            \addplot[ybar,fill=lblue,draw=blue,error bars/.cd,y dir=both,y explicit] coordinates {
                (1,   33.68) +- (0, 0.78)
                (2,  29.87)  +- (0, 0.86)
                (3,  29.83)  +- (0, 0.88)
                (4,  6.96) +- (0, 2.6)
                (5,  4.19) +- (0, 2.11)
                (6,  3.75) +- (0, 2.11)
                (7,  6.1) +- (0, 1.4)
                (8,  5.36) +- (0, 2.33)
                (9,  4.29) +- (0, 2.08)
                };
  
  \end{axis}
\end{tikzpicture}
\caption{The 9 cache sets that are used by a \SI{4096}{b} key and their error rate when recovering the key from a single trace.}
\label{fig:set-error}
\end{figure}

Figure~\ref{fig:attack-times} gives an overview of how long the individual steps of an average attack take. 
The runtime of automatic cache set detection varies depending on which cache sets are used by the victim. The attacked buffer spans $9$ cache sets, out of which $6$ show low bit-error rate, as shown in Figure~\ref{fig:set-error}. For the attack we select one of the $6$ sets, as the other $3$ suffer from too much noise. 
The noise is mainly due to the buffer not being aligned to the cache set. Furthermore, as already known from previous attacks, the hardware prefetcher can induce a significant amount of noise~\cite{Yarom2014,Gruss2015Template}.

Detecting one vulnerable cache set within all $2048$ cache sets requires about $340$ trials on average. With a monitoring time of \SI{0.21}{s} per cache set, we require a maximum of \SI{72}{s} to eventually capture a trace from a vulnerable cache set. 
Thus, based on our experiments, we estimate that cache set detection---if successful---always takes less than \SI{3}{min}.

One trace spans $220.47$ million CPU cycles on average.
Typically, \zero and \one bits are uniformly distributed in the key. 
The estimated number of multiplications is therefore half the bit size of the key. 
Thus, the average multiplication takes \SI{107662}{cycles}.
This differs from the values shown in Table~\ref{tab:keysizes} because the attacker constantly evicts the victim's buffer, inherently causing a slowdown.
In addition, one could artificially slow down a victim through constant eviction to improve the performance of cache attacks.
This is known as performance degradation~\cite{Allan2015}.
However, as the \PrimeProbe measurement takes on average \SI{734}{cycles}, we do not have to artificially slow down the victim and thus remain stealthy.

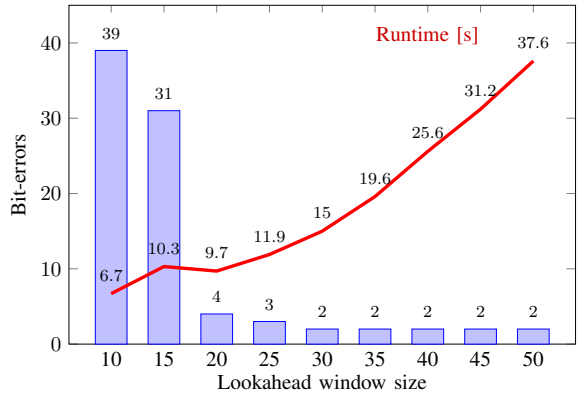
\begin{figure}[t]
\centering
\begin{tikzpicture}[scale=0.8]
  \begin{axis}[
            legend style={at={(0.5,-0.45)},
            anchor=north,legend columns=-1},
            xticklabels={10,15,20,25,30,35,40,45,50},
            xticklabel style={text height=1.5ex},
            ylabel={Bit-errors},
            xlabel={Lookahead window size},
            xtick={10,15,20,25,30,35,40,45,50},
            yscale=0.8,
            width=10cm,
            nodes near coords,
            nodes near coords align={vertical},
            enlarge x limits=0.1,
            ymin=0,
            ymax=45,
            bar width=15pt,
            every node near coord/.append style={font=\footnotesize,yshift=2pt}
          ]
            \addplot[ybar,fill=lblue!80,draw=blue] coordinates {
            
				(10,39)
				(15,31)
				(20,4)
				(25,3)
				(30,2)
				(35,2)
				(40,2)
				(45,2)
				(50,2)
                };
            \addplot[draw=red,ultra thick,yshift=0pt,yscale=1] coordinates {
				(10,6.7)
				(15,10.3)
				(20,9.7)
				(25,11.9)
				(30,15.0)
				(35,19.6)
				(40,25.6)
				(45,31.2)
				(50,37.6)
				} node[pos=1] (endofplotsquare) {};
            \node [below,yshift=20pt,xshift=-50pt,color=red!80!black] at (endofplotsquare) {Runtime [s]};

  \end{axis}
\end{tikzpicture}
\caption{Up to a lookahead window size of $30$, an increased window size reduces the number of bit errors while increasing recovery runtime. The measurement is conducted with 7 traces.}
\label{fig:traces-lookahead}
\end{figure}

When looking at a single trace, we can already recover about \SI{96}{\percent} of the RSA private key, as shown in Figure~\ref{fig:set-error}.
For a full key recovery we combine multiple traces using our key recovery algorithm, as explained in Section~\ref{sec:keyrecovery}.
We first determine a reasonable lookahead window size.
Figure~\ref{fig:traces-lookahead} shows the performance of our key recovery algorithm for varying lookahead window sizes on $7$ traces.
For lookahead windows smaller than $20$, bit errors are pretty high.
In that case, the lookahead window is too small to account for all insertion and deletion errors, causing relative shifts between the partial keys.
The key recovery algorithm is unable to align partial keys correctly and incurs many wrong ``correction'' steps, increasing the overall runtime as compared to a window size of $20$.
While a lookahead window size of $20$ already shows a good performance, a window size of $30$ or more does not significantly reduce the bit errors.
Therefore, we fixed the lookahead window size to $20$.

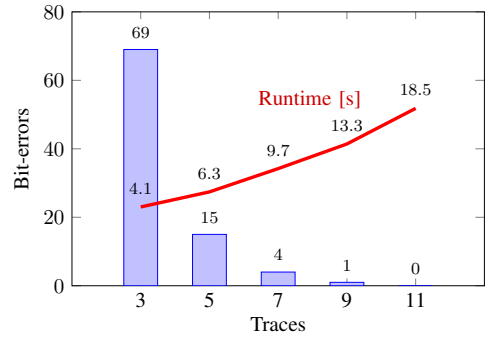
\begin{figure}[t]
\centering
\begin{tikzpicture}[scale=0.8]
  \begin{axis}[
            legend style={at={(0.5,-0.45)},
            anchor=north,legend columns=-1},
            xticklabels={3,5,7,9,11},
            xticklabel style={text height=1.5ex},
            ylabel={Bit-errors},
            xlabel={Traces},
            xtick={3,5,7,9,11},
            yscale=0.8,
            nodes near coords,
            nodes near coords align={vertical},
            enlarge x limits=0.25,
            ymin=0,
            ymax=80,
            bar width=16pt,
            every node near coord/.append style={font=\footnotesize,yshift=2pt}
          ]
            \addplot[ybar,fill=lblue!80,draw=blue] coordinates {
                (3,  69)
                (5,  15)
                (7,   4)
                (9,   1)
                (11,  0)
                };
            \addplot[draw=red,ultra thick,yshift=30pt,yscale=2.0] coordinates {
                (3,  4.1)
                (5,  6.3)
                (7,   9.7)
                (9,   13.3)
                (11,  18.5)
                } node[pos=1] (endofplotsquare) {};
            \node [below,yshift=12pt,xshift=-50pt,color=red!80!black] at (endofplotsquare) {Runtime [s]};

  \end{axis}
\end{tikzpicture}
\caption{With the number of captured traces, the number of bit errors decrease while the runtime to recover the key increases.}
\label{fig:traces-err-time}
\end{figure}

To remove the remaining bit errors and get full key recovery, we have to combine more traces. 
Figure~\ref{fig:traces-err-time} shows how the number of traces affects the key recovery performance.
We can recover the full RSA private key without any bit errors by combining only $11$ traces within just \SI{18.5}{sec}.
This results in a total runtime of less than \SI{130}{sec} for the offline key recovery process.

\subsubsection{Generalization}
Based on our experiments we can deduce that the same attacks are also possible in a weaker scenario, where only the attacker is inside the enclave.
On most computers, applications handling cryptographic keys are not protected by SGX enclaves. 
From the attacker's point of view, attacking such an unprotected application does not differ from attacking an enclave. 
We only rely on the last-level cache, which is shared among all applications, independently of whether they run inside an enclave or not. 
We empirically verified that such attacks on the outside world are possible and were again able to recover RSA private keys.

        \subsection{Virtualized Environment}
We now show that the attack also works in a virtualized environment. 

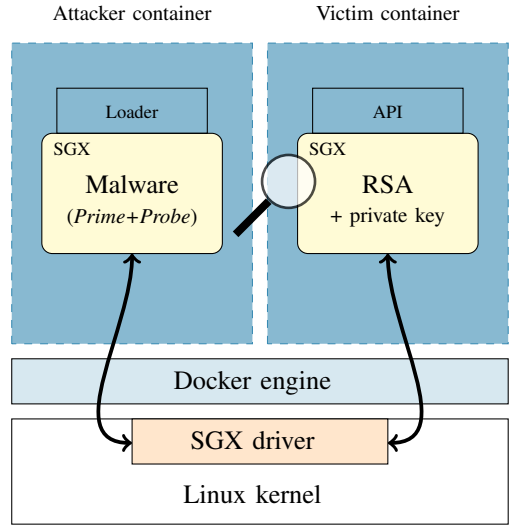
\begin{figure}
\centering

\begin{tikzpicture}[scale=0.8]
\definecolor{dockerblue}{HTML}{398bb3}

\draw (0,1) [draw=dockerblue,fill=dockerblue!60,dashed] rectangle (4,6);
\begin{scope}[shift={(0,-1)}]
\draw (0.5,3.5) [rounded corners,fill=yellow!20] rectangle (3.5,5.5);
\node at (1,5.25) {\scriptsize SGX};
\node [text centered] at (2, 4.65) {Malware};
\node [text centered] at (2, 4.1) {\footnotesize (\PrimeProbe)};

\draw (0.75,5.5) rectangle (3.25,6.25);
\node at (2,5.85) {\scriptsize Loader};
\end{scope}

\node [text centered] at (2, 6.5) {\footnotesize Attacker container};

\begin{scope}[shift={(4.25,0)}]
\draw (0,1) [draw=dockerblue,fill=dockerblue!60,dashed] rectangle (4,6);
\begin{scope}[shift={(0,-1)}]
\draw (0.5,3.5) [rounded corners,fill=yellow!20] rectangle (3.5,5.5);
\node at (1,5.25) {\scriptsize SGX};
\node [text centered] at (2, 4.65) {RSA};
\node [text centered] at (2, 4.1) {\footnotesize + private key};

\draw (0.75,5.5) rectangle (3.25,6.25);
\node at (2,5.85) {\scriptsize API};
\end{scope}

\node [text centered] at (2, 6.5) {\footnotesize Victim container};

\end{scope}

\begin{scope}[shift={(0,-0.75)}]
\draw (0,-1.25) rectangle (8.25,0.5);
\node [text centered] at (4, -0.75) {Linux kernel};

\draw (2,-0.25) [fill=orange!20] rectangle (6.25,0.5);
\node [text centered] at (4, 0.15) {SGX driver};
\end{scope}

\draw (0,-0) [fill=dockerblue!20] rectangle (8.25,0.75);
\node [text centered] at (4, 0.35) {Docker engine};

\draw (2,2.5) edge[out=270,in=180,<->,line width=0.5mm] (2,-0.65);
\draw (6.25,2.5) edge[out=270,in=0,<->,line width=0.5mm] (6.25,-0.65);

\begin{scope}[shift={(14.5,1.5)},xscale=-1,scale=1.1]
\draw [draw=black,line width=3pt] (9.3,1.7) to (9.8,1.2);
\draw [draw=black,line width=1pt,fill=white!10,opacity=0.7] (9,2) ellipse (0.4cm and 0.4cm);
\end{scope}
\end{tikzpicture}
\caption{Running the SGX enclaves inside Docker containers to provide further isolation. The host provides both containers access to the same SGX driver.}
\label{fig:docker}
\end{figure}

As described in Section~\ref{sec:sgx}, no hypervisor with SGX support was available at the time of our experiments. 
Instead of full virtualization using a virtual machine, we used the lightweight Docker containers.
Docker containers are also used by large cloud providers, \eg Amazon~\cite{Docker_AWS} or Microsoft Azure~\cite{Microsoft_Docker}. 
To enable SGX within a container, the host operating system has to provide SGX support. 
The SGX driver is then simply shared among all containers. 
Figure~\ref{fig:docker} shows our setup where the SGX enclaves communicate directly with the SGX driver of the host operating system.
Applications running inside the container do not experience any difference to running on a native system. 
They can use any functionality provided by the host operating system.
Consequently, the unmodified malware also works inside containers.

Considering the performance within Docker, only I/O operations and network access have a measurable overhead~\cite{Felter2015}.
Operations that only depend on memory and CPU do not see any performance penalty, as these operations are not virtualized.
Thus, caches are also not affected by the container. 

We were successfully able to attack a victim from within a Docker container without any changes in the malware. 
We can even perform a cross-container attack, \ie both the malware and the victim are running inside different containers, without any changes. 
As expected, we require the same number of traces for a full key recovery. 
These results confirm that containers do not provide additional protection against our malware at all. 

Furthermore, we can speculate whether our malware would also work within virtual machines based on the experimental KVM support description~\cite{Intel2016kvmsgx}.
Many cross-VM cache attacks have been demonstrated in the past years~\cite{Apecechea2014a,Inci2015,Liu2015}, as the CPU cache is a shared resource in virtual machines.
This does not change with SGX, and thus, enclaves inside virtual machines will also share the last-level cache. 
The experimental implementation for KVM relies on the host system's SGX driver to provide memory pages to the enclave inside the virtual machine.
We thus expect that our malware will work across virtual machines either with only minor changes, or even without any adaptations.
        
%
\section{Countermeasures} \label{sec:countermeasures}
In this section, we discuss advantages and disadvantages of different countermeasures. 
Previously presented countermeasures mostly cannot be applied to a scenario where a malicious enclave
performs a cache attack and no assumptions about the operating system are made.
We group countermeasures into 3 categories, based on whether they require:
\begin{compactenum}
\item a modification of the enclave (source level),
\item a modification of the operating system (OS level) assuming the operating system is benign,
\item a change in hardware (hardware level).
\end{compactenum}

\subsection{Source Level}
\subsubsection{Exponent Blinding}
A generic side-channel protection for RSA is exponent blinding~\cite{Kocher1996}. 
To sign a message $m$, the signer generates a random blinding value $k$ for each signature. 
The signer then calculates the signature as $m^{d + k \cdot \phi(N)} \mod N$ where $d$ is the private key and $N$ is the RSA modulus. 

An attacker will only be able to measure the blinded exponent on every execution. 
When a single-trace key recovery is not possible, the attacker has to wait for collisions, \ie signatures where the same blinding was used. 
For a sufficiently large blinding factor $k$, \eg \SI{64}{bit}, this becomes infeasible in practice. 
As the exponent grows with the blinding factor, this solution is a trade-off between performance and side-channel resistance. 
This has no effect if key recovery from a single trace is possible, only if more than one trace is required.
Furthermore, this countermeasure relies on the presence of a random number source. 

Exponent blinding is specific to certain cryptographic operations, such as RSA signature computations. 
It will prevent the proposed attack, but other parts of the signature process might still be vulnerable to an attack~\cite{Schindler2015}.

\subsubsection{Bit Slicing}
Bit slicing is a technique originally proposed by Biham~\cite{Biham1997} to improve the performance of DES.
Matsui~\cite{Matsui2006} was the first to show a bit-sliced implementation of AES. 
Sudhakar~\etal\cite{Sudhakar2007} presented a bit-sliced Montgomery multiplication for RSA and ECC. 
The main idea of bit slicing is to use only bit operations for computations throughout the algorithm. 
No lookup tables or branches are used in these algorithms and thus, they are not vulnerable to cache attacks. 

Again, this countermeasure is specific to certain cryptographic algorithms.
It requires the support of the used cryptography library and hardware support for streaming SIMD (SSE) instructions is necessary to achieve a reasonable performance~\cite{Kasper2009}. 
Bit slicing can be a good software solution while there is no hardware countermeasure. 
Other countermeasures for cryptographic implementations have been discussed by Ge~\etal\cite{Ge2016}.

\subsection{Operating System Level}
Implementing countermeasures against malicious enclave attacks on the operating system level requires trusting the operating system.
This would weaken the trust model of SGX enclaves significantly and is thus unrealistic. 
However, we want to discuss the different possibilities, in order to provide valuable information for the design process of future enclave systems.

\subsubsection{Eliminating Timers}
Removing access to high-resolution timers~\cite{Percival2005,Gullasch2011} or decreasing the accuracy~\cite{Hu1992fuzzy,Vattikonda2011,Martin2012} is often discussed as a countermeasure against cache attacks.
However, our results using the timing counter show that removing precise timers is not a viable countermeasure, as we are still able to mount a high-resolution \PrimeProbe attack. 
Moreover, on recent microarchitectures, we can even get a higher resolution using our timing thread than with the native high-resolution timestamp counter. 

However, it is possible to remove access to high-resolution timers and all forms of simultaneous multithreading to prevent this alternative approach. This would effectively eliminate access to sufficiently accurate timers and mitigate many attacks.

\subsubsection{Detecting Malware}
One of the core ideas of SGX is to remove the cloud provider from the root of trust.
If the enclave is encrypted and only decrypted after successful remote attestation, the cloud provider has no way to access the secret code inside the enclave. 
However, eliminating this core feature of SGX could mitigate malicious enclaves in practice as the enclave binary or source code could be read by the cloud provider and scanned for malicious activities.

Heuristic methods, such as behavior-based detection, are not applicable, as the malicious enclave does not rely on API calls or user interaction. 
Furthermore, for encrypted enclave code, a signature-based virus scanner has no access to the code, and the malware can easily change its signature by either re-encryption or modification of the plaintext. 
Thus, only the host binary---which contains no malicious code---can be inspected by a virus scanner. 

\begin{figure}
\centering
\begin{tikzpicture}[scale=0.8]
  \begin{axis}[
	    ybar,
	    legend style={at={(0.5,-0.3)},
	    anchor=north,legend columns=-1,draw=none},
        symbolic x coords={L1 Hits, L1 Misses, L3 Hits, L3 Misses},
	    ylabel={Performance counter value},
	    width=\hsize,
	    height=5.5cm,
        xtick=data,
        yscale=0.8,
        ymin=0,
        enlarge x limits={abs=0.8cm},
          ]
	  \addplot coordinates {
                (L1 Hits,  44880486)
                (L1 Misses,  7273437)
                (L3 Hits,   17579294)
                (L3 Misses,  4222946)
                };
	  \addplot coordinates {
                (L1 Hits, 47804802)
                (L1 Misses,  8907188)
                (L3 Hits,   18314812)
                (L3 Misses,  4418656)
            };
          \addplot coordinates {
                (L1 Hits, 1472836926)
                (L1 Misses,  1136066543)
                (L3 Hits,   532820781)
                (L3 Misses,  524571312)
            };
            
	  \legend{Debug,Pre-Release,Native}

  \end{axis}
\end{tikzpicture}
\caption{Performance counters for caches are disabled in an enclave. \FlushReload of one variable in a loop results in a high cache activity, which can be seen in native environment, but not on SGX debug or pre-release mode.}
\label{fig:perf-counter}
\end{figure}
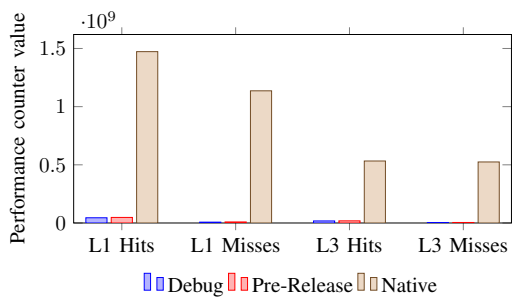

Herath and Fogh~\cite{Herath2015} proposed to use hardware performance counters to detect cache attacks. Subsequently, several other approaches instrumenting performance counters to detect cache attacks have been proposed~\cite{Chiappetta2015,Gruss2016Flush,Payer2016}. However, according to Intel, SGX enclave activity is not visible in the thread-specific performance counters~\cite{Intel_SGXDifferences}. 
We verified that even performance counters for last-level cache accesses are disabled for enclaves. 
Figure~\ref{fig:perf-counter} shows the results of a simple test program running inside a debug and pre-release enclave, and without an enclave. 
The visible cache hits and misses are caused by the host application only.
This makes it impossible for current anti-virus software and other detection mechanisms to detect the malware.

\subsubsection{Enclave Coloring}
We propose enclave coloring as an effective countermeasure against cross-enclave attacks. 
Enclave coloring is a software approach to partition the cache into multiple smaller parts. 
Each of the parts spans over multiple cache sets, and no cache set is included in more than one part. 
An enclave gets one or more such cache parts. 
This assignment of cache parts is either done by the hardware or by a trusted operating system. 

If implemented in software, the operating system can split the last-level cache through memory allocation.
The cache set is determined by bits of the physical address. 
The lower bits of the cache set index are below bit $12$ and therefore determined by the page offset, \ie the data's position within a \SI{4}{\KB} page. 
The upper bits of the cache set are not visible to the enclave application and can thus be controlled by the operating system when allocating pages. 
We call these upper bits a color. 
Whenever an enclave requests pages from the operating system (we consider the SGX driver as part of the operating system), it will only get pages with a color that is not present in any other enclave. 
This coloring ensures, that two enclaves cannot have data in the same cache set, and thus an eviction of the data---and therefore a \PrimeProbe attack---is not possible across enclaves. 
However, attacks on the operating system or other processes on the same host would still be possible.

Enclave coloring requires a trusted operating system, and is therefore not always applicable as it contradicts SGX's idea of having an untrusted operating system~\cite{Costan2016}. 
If the operating is trusted, this is an effective countermeasure against cross-enclave cache attacks. 

To prevent attacks on the operating system or other processes, it would be necessary to partition the rest of the memory as well, \ie system-wide cache coloring~\cite{Raj2009}.
Godfrey~\etal\cite{Godfrey2014} evaluated a coloring method for hypervisors by assigning every virtual machine a partition of the cache. 
They concluded that this method is only feasible for a small number of partitions. 
As the number of simultaneous enclaves is relatively limited by the available amount of SGX memory, enclave coloring can be applied to prevent cross-enclave attacks. 
Protecting enclaves from malicious applications or preventing malware inside enclaves is however not feasible using this method.

\subsubsection{Heap Randomization}
Our attack relies on the fact, that the used buffers for the multiplication are always at the same memory location. 
This is indeed the case, as the memory allocator (\texttt{dlmalloc}) has a deterministic behavior and uses a best-fit approach for moderate buffer sizes as used in the RSA implementation. 
Freeing a buffer and allocating it again will always result in the same memory location for the buffer. 

We suggest randomizing the heap allocations for security relevant data such as the used buffers. 
A randomization of the addresses and thus cache sets bears two advantages. 
First, an automatic cache set detection is not possible anymore, as the identified set will change for the next run of the algorithm. 
Second, if more than one trace is required to reconstruct the key, this countermeasure increases the number of required traces by multiple orders of magnitude as the probability to measure the correct cache set decreases. 

Although not obvious at first glance, this method requires a certain amount of trust in the underlying operating system. 
A malicious operating system could assign only pages mapping to certain cache sets to the enclave, similar to enclave coloring. 
Thus, the randomization is limited to only a subset of cache sets, increasing the probability for an attacker to measure the correct cache set from \SI{0.1}{\percent} to \SI{7}{\percent}.

\subsubsection{Intel CAT}
Recently, Intel introduced an instruction set extension called CAT (cache allocation technology)~\cite{Intel_vol3}. With Intel CAT it is possible to restrict CPU cores to one of the slices of the last-level cache and even to pin cache lines. Liu~\etal\cite{Liu2016catalyst} proposed a system that uses CAT to protect general purpose software and cryptographic algorithms. Their approach can be directly applied to protect against a malicious enclave. However, this approach also does not allow to protect enclaves from an outside attacker.

\subsection{Hardware Level}
\subsubsection{Combining Intel CAT with SGX}
Instead of using Intel CAT on the operating level it could also be used to protect enclaves on the hardware level. By changing the \texttt{eenter} instruction in a way that it implicitly activates CAT for this core, any cache sharing between SGX enclaves and the outside as well as co-located enclaves could be eliminated. Thus, SGX enclaves would be protected from outside attackers. Furthermore, it would protect co-located enclaves as well as the operating system and user programs against malicious enclaves.

\subsubsection{Secure RAM}
To fully mitigate cache- or DRAM-based side-channel attacks memory must not be shared among processes. 
We propose an additional secure memory element that resides inside the CPU. 
Data stored within this memory is not cachable, thus the memory has to be fast to not incur performance penalties. 

The SGX driver can then provide a special API to acquire this element for temporarily storing sensitive data. 
A cryptographic library could use this memory to execute code which depends on secret keys such as the square-and-multiply algorithm. 
Providing such a secure memory element per CPU core would even allow parallel execution of multiple enclaves. 

As data from this element is only accessed by one program and is never cached, cache attacks and DRAM-based attacks are not possible anymore. 
Moreover, if this secure memory is inside the CPU, it is infeasible for an attacker to mount physical attacks or to probe the memory bus. 
It is unclear whether Intel's eDRAM implementation can already be abused as a secure memory to protect applications against cache attacks. 

\section{Conclusion}\label{sec:conclusion}
There have been speculations that SGX could be vulnerable to cache side-channel attacks and might allow the implementation of super malware.
However, Intel claimed that SGX features impair side-channel attacks and recommends using SGX enclaves to protect cryptographic computations.
Furthermore, it was presumed that they cannot perform harmful operations.

In this paper, we demonstrated the first malware running in real SGX hardware enclaves.
We demonstrated private key theft in a fully automated end-to-end attack from a co-located SGX enclave, despite all restrictions of SGX, \eg no timers, no large pages, no physical addresses, and no shared memory.

We developed the most accurate timing measurement technique currently known for Intel CPUs, perfectly tailored to the hardware.
We combined DRAM and cache side channels, to build a novel approach that recovers physical address bits without assumptions on the page size.
We attack the RSA implementation of \mbedTLS that is used for instance in OpenVPN.
The attack succeeds despite protection against side-channel attacks using a constant-time multiplication primitive.
We extract \SI{96}{\percent} of a 4096-bit RSA private key from a single \PrimeProbe trace and achieve full key recovery from only 11 traces within 5 minutes.

Besides not fully preventing malicious enclaves, SGX provides protection features to conceal attack code.
Even the most advanced detection mechanisms using performance counters cannot detect our malware.
Intel intentionally does not include SGX activity in the performance counters for security reasons.
However, this unavoidably provides attackers with the ability to hide attacks as it eliminates the only known technique to detect cache side-channel attacks.
We discussed multiple design issues in SGX and proposed countermeasures that should be considered for future versions.


\section*{Acknowledgments}
This project has received funding from the European Research Council (ERC) under the European Union’s Horizon 2020 research and innovation programme  
\noindent\begin{tabular}{m{\dimexpr 0.13\hsize} m{8pt} m{\dimexpr 0.13\hsize} m{2pt} m{\dimexpr 0.88\hsize-16\tabcolsep-2pt}}
 \vspace*{1mm}\includegraphics[height=1.1cm]{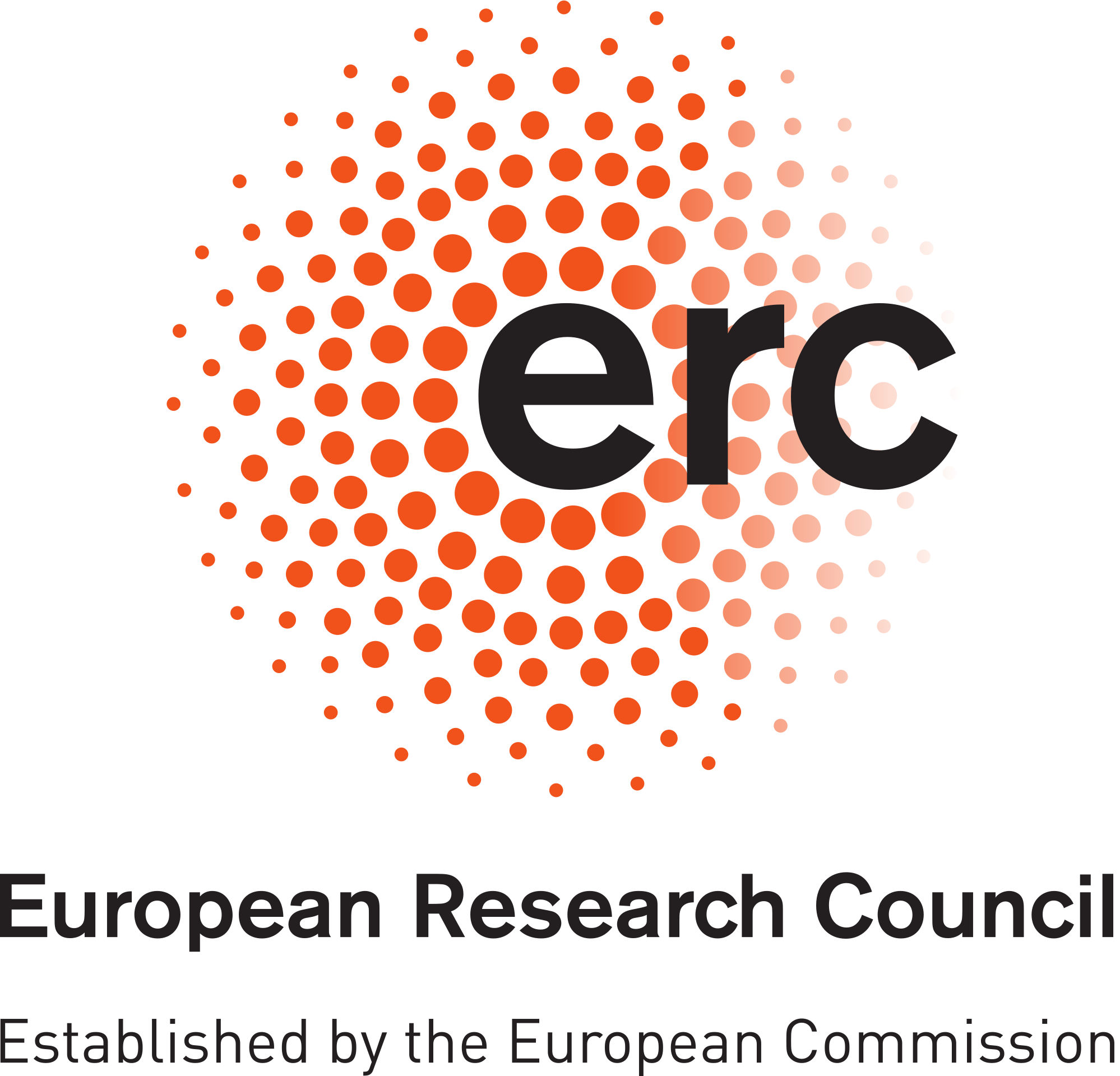}& & \vspace*{2mm}\includegraphics[height=1.1cm]{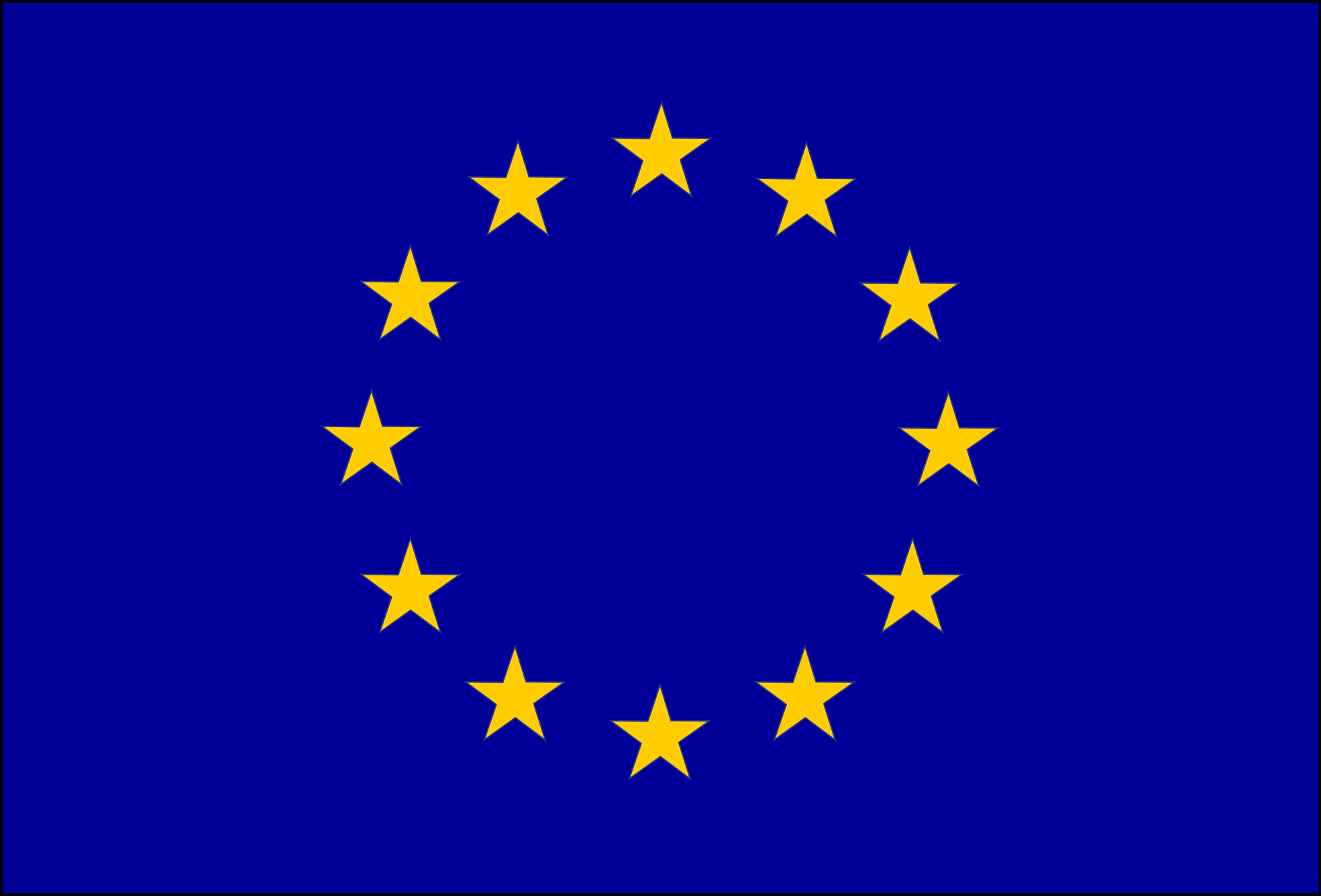} & &  (grant agreement No 681402). This work was partially supported by the TU Graz LEAD project "Dependable Internet of Things in Adverse Environments".
 \quad \quad \quad \vspace{-\baselineskip} 
\end{tabular} \\


\bibliographystyle{IEEEtran}
\bibliography{IEEEabrv,bibliography}

\end{document}